\documentclass{aa}

\usepackage{natbib}
\bibpunct{(}{)}{;}{a}{}{,} 

\usepackage{float}
\usepackage{caption}
\DeclareCaptionFormat{cont}{#1 (cont.)#2#3\par} 
\usepackage{txfonts}
\usepackage{siunitx}  
\usepackage{color}
\usepackage[titletoc,title]{appendix}

\newcommand{\tablenote}[1]{\parbox{18.3cm}{\indent \footnotesize{#1}}}

\definecolor{pink}{rgb}{0.858, 0.188, 0.478}
\definecolor{purple}{rgb}{0.53, 0.15, 0.34}

\begin{document}

\title{ Abundance of SiC$_2$ in carbon star envelopes\thanks{Based on observations carried out with the IRAM 30m Telescope. IRAM is supported by INSU/CNRS (France), MPG (Germany), and IGN (Spain).}}

\subtitle{Evidence that SiC$_2$ is a gas-phase precursor of SiC dust}

\titlerunning{SiC$_2$ in carbon star envelopes}
\authorrunning{Massalkhi et al.}

\author{
Sarah~Massalkhi\inst{1}, M.~Ag\'undez\inst{1}, J.~Cernicharo\inst{1}, L. Velilla Prieto\inst{1}, \\J. R. Goicoechea\inst{1}, G. Quintana-Lacaci\inst{1}, J. P. Fonfr\'ia\inst{1}, J. Alcolea\inst{2}, and V. Bujarrabal\inst{3}
}

\institute{
Grupo de Astrof\'isica Molecular, Instituto de Ciencia de Materiales de Madrid, CSIC, C/ Sor Juana In\'es de la Cruz 3, \\ 28049, Cantoblanco, Spain \and
Observatorio Astron\'omico Nacional (IGN), C/ Alfonso XII 3, 28014, Madrid, Spain \and
Observatorio Astron\'omico Nacional (IGN), Apartado de Correos 112, 28803, Alcal\'a de Henares, Madrid, Spain
}

\date{Received; accepted}

\abstract
{Silicon carbide dust is ubiquitous in circumstellar envelopes around C-rich asymptotic giant branch (AGB) stars. However, the main gas-phase precursors leading to the formation of SiC dust have not yet been identified. The most obvious candidates among the molecules containing an Si--C bond detected in C-rich AGB stars are SiC$_2$, SiC, and Si$_2$C. To date, the ring molecule SiC$_2$ has been observed in a handful of evolved stars, while SiC and Si$_2$C have only been detected in the C-star envelope IRC\,+10216.}
{We aim to study how widespread and abundant SiC$_2$, SiC, and Si$_2$C are in envelopes around C-rich AGB stars, and whether or not these species play an active role as gas-phase precursors of silicon carbide dust in the ejecta of carbon stars.}
{We carried out sensitive observations with the IRAM 30m telescope of a sample of 25 C-rich AGB stars to search for emission lines of SiC$_2$, SiC, and Si$_2$C in the $\lambda$ 2 mm band. We performed non-LTE excitation and radiative transfer calculations based on the LVG method to model the observed lines of SiC$_2$ and to derive SiC$_2$ fractional abundances in the observed envelopes.}
{We detect SiC$_{2}$ in most of the sources, SiC in about half of them, and do not detect Si$_2$C in any source except  IRC\,+10216. Most of these detections are reported for the first time in this work. We find a positive correlation between the SiC and SiC$_2$ line emission, which suggests that both species are chemically linked; the SiC radical is probably  the photodissociation product of SiC$_2$ in the external layer of the envelope. We find a clear trend where the denser the envelope, the less abundant SiC$_2$ is. The observed trend is interpreted as  evidence of efficient incorporation of SiC$_2$ onto dust grains, a process that is favored at high densities owing to the higher rate at which collisions between particles take place.}
{The observed behavior of a decline in the SiC$_2$ abundance with increasing density strongly suggests that SiC$_{2}$ is an important gas-phase precursor of SiC dust in envelopes around carbon stars.}

\maketitle

\section{Introduction}

\begin{table*}
\caption{Sample of carbon stars}\label{table:sources}
\centering
\resizebox{\linewidth}{!}{
\begin{tabular}{lrrcrcrccccc}
\hline \hline
\multicolumn{1}{l}{Name} & \multicolumn{1}{c}{R.A.} & \multicolumn{1}{c}{Dec.} & \multicolumn{1}{c}{$V_{\rm LSR}$} & \multicolumn{1}{c}{$D$} & \multicolumn{1}{c}{$T_{\star}$} & \multicolumn{1}{c}{$L_{\star}$} & \multicolumn{1}{c}{$\dot{M}$} & \multicolumn{1}{c}{$V_{\rm exp}$} &   \multicolumn{1}{c}{$T_{\rm d}(r_c)$} & \multicolumn{1}{c}{$r_c$} & $\Psi$ \\

\multicolumn{1}{c}{}          & \multicolumn{1}{c}{J2000.0} & \multicolumn{1}{c}{J2000.0} & \multicolumn{1}{c}{(km~s$^{-1}$)} & \multicolumn{1}{c}{(pc)} & \multicolumn{1}{c}{(K)} & \multicolumn{1}{c}{(L$_{\odot}$)}  & \multicolumn{1}{c}{(M$_{\odot}$ yr$^{-1}$)} & \multicolumn{1}{c}{(km~s$^{-1}$)} &\multicolumn{1}{c}{(K)} & \multicolumn {1}{c}{(cm)} &   \\
\hline
IRC\,+10216 & 09:47:57.45 & $+$13:16:43.9 & $-26.5$ & 130 $^{[a]}$  & 2330 $^{[a]}$  &  8750 $^{[a]}$  & $2.0\times10^{-5}$ $^{[a]}$  & 14.5 & 800 $^{[a]}$  &  $2.0\times10^{14}$ $^{[a]}$ &  300 $^{[a]}$   \\
CIT\,6      & 10:16:02.27 & $+$30:34:18.6 & $-1$    & 440 $^{[g]}$ & 1800 $^{[g]}$ & 10000 $^{[g]}$ & $6.0\times10^{-6}$ $^{[g]}$  & 17   & 1000 $^{[g]}$ & $2.1\times10^{14}$ $^{[g]}$ & 141 $^{[d]}$    \\
CRL\,3068   & 23:19:12.24 & $+$17:11:33.4 & $-31.5$  & 1300 $^{[g]}$ & 1800 $^{[g]}$  & 10900 $^{[g]}$ & $2.5\times10^{-5}$ $^{[g]}$  & 14.5  & 1500 $^{[g]}$      & $2.0\times10^{14}$ $^{[g]}$  & 174 $^{[d]}$ \\ 
S\,Cep      & 21:35:12.83 & $+$78:37:28.2 & $-15.3$     & 380 $^{[k]}$    & 2200 $^{[k]}$   & 7300 $^{[k]}$  & $1.2\times10^{-6}$ $^{[k]}$  & 22.5 & 1400 $^{[k]}$    & $5.8\times10^{13}$ $^{[k]}$ & 360 $^{[d]}$ \\
IRC\,+30374 & 19:34:09.87 & $+$28:04:06.3 & $-12.5$     & 1200 $^{[j]}$   & 2000 $^{[j]}$   & 9800 $^{[j]}$   & $1.0\times10^{-5}$ $^{[j]}$     & 25 & 1000 $^{[j]}$    & $2.2\times10^{14}$ $^{[j]}$ & 1008 $^{[d]}$ \\
Y\,CVn      & 12:45:07.83 & $+$45:26:24.9 & $+22$       & 220 $^{[k]}$    & 2200 $^{[k]}$   & 4400 $^{[k]}$ & $1.5\times10^{-7}$ $^{[k]}$   & 7  & 1500 $^{[k]}$    & $8.7\times10^{13}$ $^{[k]}$ & 500 $^{[h]}$  \\
LP\,And     & 23:34:27.53 & $+$43:33:01.2 & $-17$       & 630 $^{[g]}$    & 1900 $^{[g]}$   & 9600 $^{[g]}$  & $7.0\times10^{-6}$ $^{[g]}$ &14.5  & 1100 $^{[g]}$    & $1.8\times10^{14}$ $^{[g]}$ & 288 $^{[d]}$  \\ 
V\,Cyg      & 20:41:18.27 & $+$48:08:28.8 & $+13.5$     & 366 $^{[g]}$    & 2300 $^{[g]}$   & 6000 $^{[g]}$  & $1.6\times10^{-6}$ $^{[g]}$   & 12 & 1400 $^{[g]}$   & $9.4\times10^{13}$ $^{[g]}$ & 364 $^{[d]}$   \\
UU\,Aur     & 06:36:32.84 & $+$38:26:43.8 & $+6.7$  & 260 $^{[k]}$   & 2800 $^{[k]}$   & 6900 $^{[k]}$ & $2.4\times10^{-7}$ $^{[k]}$   & 10.6    & 1500 $^{[k]}$   & $6.3\times10^{13}$ $^{[k]}$ & 11111 $^{[b]}$\\ 
V384\,Per    & 03:26:29.51 & $+$47:31:48.6 & $-16.8$     & 560 $^{[k]}$   & 2000 $^{[k]}$   & 8100 $^{[k]}$   & $2.3\times10^{-6}$ $^{[b]}$  & 15.5 & 1300 $^{[k]}$    & $1.0\times10^{14}$ $^{[k]}$ & 584 $^{[d]}$  \\
IRC\,+60144 & 04:35:17.54 & $+$62:16:23.8 & $-48.8$     & 1030 $^{[b]}$  & 2000 $^{[b]}$ & 7800 $^{[b]}$   & $3.7\times10^{-6}$ $^{[b]}$   & 19.5 & 1200 $^{[n]}$    & $2.0\times10^{14}$  $^{[b]}$  & 11 $^{[b]}$\\
U\,Cam      & 03:41:48.17 & $+$62:38:54.4 & $+6$        & 430 $^{[i]}$    & 2695 $^{[i]}$  & 7000 $^{[i]}$ & $2.0\times10^{-7}$ $^{[i]}$  & 13 & 1500 $^{[j]}$   & $4.4\times10^{13}$ $^{[j]}$ & 833 $^{[i]}$  \\  
V636\,Mon   & 06:25:01.43 & $-$09:07:15.9 & $+10$       & 880 $^{[f]}$   & 2500 $^{[n]}$   & 8472 $^{[e]}$  & $5.8\times10^{-6}$ $^{[f]}$    & 20 & 1200 $^{[n]}$ & $1.7\times10^{14}$ $^{[n]}$   & 300 $^{[n]}$ \\
IRC\,+20370 & 18:41:54.39 & $+$17:41:08.5 & $-0.8$      & 600 $^{[k]}$    & 2200 $^{[k]}$   & 7900 $^{[k]}$   & $3.0\times10^{-6}$  $^{[b]}$     & 14& 1500 $^{[k]}$    & $8.1\times10^{13}$ $^{[k]}$ & 266 $^{[d]}$ \\
R\,Lep      & 04:59:36.35 & $-$14:48:22.5 & $+11.5$     & 432 $^{[b]}$   & 2200 $^{[b]}$  & 5500 $^{[b]}$  & $8.7\times10^{-7}$ $^{[b]}$   & 17.5& 1000 $^{[g]}$    & $1.8\times10^{14}$  $^{[b]}$  & 2000 $^{[b]}$\\ 
W\,Ori      & 05:05:23.72 & $+$01:10:39.5 & $-1$        & 220 $^{[k]}$    & 2600 $^{[k]}$   & 3500 $^{[k]}$   & $7.0\times10^{-8}$ $^{[k]}$     & 11& 1500 $^{[k]}$    & $4.3\times10^{13}$ $^{[k]}$ & 333 $^{[h]}$  \\
CRL\,67     & 00:27:41.10 & $+$69:38:51.5 & $-27.5$     & 1410 $^{[d]}$    & 2500 $^{[n]}$  & 9817 $^{[d]}$    & $1.1\times10^{-5}$ $^{[d]}$    & 16& 1200 $^{[n]}$   & $1.8\times10^{14}$ $^{[n]}$   & 495 $^{[d]}$\\
CRL\,190    & 01:17:51.62 & $+$67:13:55.4 & $-39.5$     & 2790 $^{[d]}$    & 2500 $^{[c]}$  & 16750 $^{[d]}$   & $6.4\times10^{-5}$ $^{[d]}$      & 17& 1000 $^{[c]}$    & $4.7\times10^{14}$  $^{[c]}$ & 424 $^{[d]}$  \\
S\,Aur      & 05:27:07.45 & $+$34:08:58.6 & $-17$       & 300 $^{[k]}$    & 3000 $^{[k]}$   & 8900 $^{[k]}$   & $4.0\times10^{-7}$ $^{[k]}$   & 24.5& 1500 $^{[k]}$    & $7.3\times10^{13}$ $^{[k]}$ & 500 $^{[h]}$  \\ 
V\,Aql      & 19:04:24.15 & $-$05:41:05.4 & $+53.5$     & 330 $^{[k]}$    & 2800 $^{[k]}$   & 6500 $^{[k]}$   & $1.4\times10^{-7}$ $^{[k]}$   & 8& 1500 $^{[k]}$    & $6.1\times10^{13}$ $^{[k]}$ & 500 $^{[h]}$   \\
CRL\,2513     & 20:09:14.25 & $+$31:25:44.9 & $+17.5$     & 1760 $^{[d]}$   & 2500  $^{[c]}$  & 8300 $^{[c]}$ & $2.0\times10^{-5}$ $^{[d]}$    & 25.5  & 1200 $^{[c]}$    & $1.6\times10^{14}$  $^{[c]}$ & 453 $^{[d]}$   \\
CRL\,2477     & 19:56:48.43 & $+$30:43:59.9 & $+5$        & 3380 $^{[m]}$   & 3000 $^{[m]}$ & 13200 $^{[m]}$  & $1.1\times10^{-4}$ $^{[m]}$    & 20 & 1800 $^{[m]}$ & $2.8\times10^{14}$ $^{[m]}$& 532 $^{[d]}$    \\
CRL\,2494     & 20:01:08.51 & $+$40:55:40.2 & $+29$       & 1480 $^{[b]}$    & 2400 $^{[b]}$    & 10200 $^{[b]}$  & $7.5\times10^{-6}$ $^{[b]}$   & 20 & 1200 $^{[n]}$        & $2.3\times10^{14}$ $^{[b]}$  & 436 $^{[d]}$  \\
RV\,Aqr     & 21:05:51.74 & $-$00:12:42.0 & $+0.5$      & 670 $^{[k]}$    & 2200 $^{[k]}$     & 6800 $^{[k]}$     & $2.3\times10^{-6}$ $^{[b]}$   & 15 & 1300 $^{[k]}$   & $7.6\times10^{13}$ $^{[k]}$ & 200 $^{[h]}$  \\
ST\,Cam     & 04:51:13.35 & $+$68:10:07.6 &  -13.6     & 360 $^{[k]}$   & 2800 $^{[k]}$    & 4400 $^{[k]}$   &  $1.3\times10^{-7}$ $^{[k]}$   & 8.9 & 1500 $^{[k]}$   & $5.0\times10^{13}$ $^{[k]}$   & 500 $^{[h]}$    \\ \hline
\end{tabular}
}
\tablenote{\\
References: $^{[a]}$~Ag\'undez~et~al.~(2012), $^{[b]}$~Danilovich~et~al.~(2015), $^{[c]}$~Groenewegen~et~al.~(1998), $^{[d]}$~Groenewegen~et~al.~(2002), $^{[e]}$~Guandalini~et~al.~(2013), $^{[f]}$~Guandalini~et~al.~(2006) $^{[g]}$~Ramstedt~et~al.~(2014), $^{[h]}$~Sch\"oier \& Olofsson~et~al.~(2001), $^{[i]}$~Parameters of the present-day wind of U\,Cam from Sch\"oier et al.~(2005), $^{[j]}$~Sch\"oier et al.~(2006), $^{[k]}$~Sch\"oier et al.~(2013), $^{[m]}$~Speck~et~al.~(2009), $^{[n]}$ Assumed value for the stellar effective temperature  $T_{\star}$ is 2500 K, for the condensation radius $r_{c}$ is 5 R$_{\star}$, for the dust temperature at the condensation radius $T_{\rm d}(r_{\rm c})$ is 1200 K, and for the gas-to-dust mass ratio $\Psi$ is 300.}
\end{table*}

Low- and intermediate-mass stars, with initial masses $<8$ M$_{\odot}$, experience in their late stages of evolution known as the  asymptotic giant branch (AGB), intense mass-loss processes which result in circumstellar envelopes (CSEs) made up of molecules and dust grains. AGB stars are the main sources of interstellar dust in the Galaxy \citep{geh1989}. The chemical nature of the synthesized dust is determined by the C/O elemental abundance ratio at the surface of the AGB star. Dust is mainly composed of silicates in envelopes around oxygen-rich stars (C/O $<$ 1), while carbonaceous and silicon carbide dust is formed around carbon-rich stars (C/O $>$ 1) (e.g., \citealt{Swa2005}). The dust formation process involves a first step in which condensation nuclei of nanometer size are formed from some gas-phase precursors of highly refractory character, a process that takes place in the surroundings of the stellar atmosphere. The second step involves the growth of the nuclei to micrometer sizes by accretion and coagulation as the material is pushed out by the stellar wind \citep{gai1984}. What  the gas-phase building blocks of condensation nuclei are and how these particles evolve toward the micrometer-sized grains that populate the interstellar medium are key questions that are not yet well understood. 

In this article we  perform an observational study to constrain  the main gas-phase precursors of silicon carbide (SiC) dust in C-rich AGB stars. The presence of SiC dust in circumstellar envelopes around C-rich AGB stars was established through the observation of a solid-state emission band at $\sim$11.3~$\mu$m \citep{hac1972,tre1974}. The 11.3~$\mu$m feature has been observed toward a large number of C-rich AGB stars with the \emph{IRAS} and \emph{ISO} satellites (e.g., \citealt{lit1986,cha1990,yan2004}). However, what  the gas-phase precursors of SiC dust are is still a pending question.

Various molecules containing an Si$-$C bond have been observed in envelopes around C-rich AGB stars and are potential gas-phase precursors of SiC dust. The ring molecule SiC$_2$ has long been observed in the atmospheres of optically visible carbon stars through the  Merrill--Sanford electronic bands (see, e.g., \citealt{sar2000}). \citet{tha1984} reported the first observation of the rotational spectrum of SiC$_2$ toward the highly reddened envelope IRC\,+10216;  since then observations of this molecule have been reported in only  a few other C-rich AGB or post-AGB objects: IRAS\,15194$-$5115 \citep{nym1993}, CRL\,2688 \citep{bac1997}, CIT\,6 \citep{zha2009a}, and CRL\,3068 \citep{zha2009b}. The diatomic molecule SiC was found by \citet{cer1989} in IRC\,+10216, and more recently \citet{cer2015} have reported on the discovery of Si$_2$C toward the same source. IRC\,+10216 is the only source where SiC$_2$ has been thoroughly studied across the mm and sub-mm ranges \citep{luc1995,cer2010,pat2011,mul2012,fon2014,vel2015} and where SiC and Si$_2$C detections have been reported. The scenario emerged from the studies of IRC\,+10216 is that only SiC$_2$ and Si$_2$C are present in the innermost regions, while SiC is probably a photodissociation product of these two triatomic molecules and it is thus present in the outer envelope. The U-shaped line profiles of SiC \citep{cer2000} and the upper limits derived to its abundance in the inner envelope \citep{vel2015} support this idea. Moreover, chemical equilibrium calculations predict abundant SiC$_2$ and Si$_2$C  but little SiC in the hot and dense surroundings of the AGB star \citep{tej1991,yas2012,cer2015}. It therefore seems that the gas-phase molecule SiC is not an important building block of SiC dust, while SiC$_2$ and Si$_2$C are probably the main gas-phase precursors. In addition to these species, other molecules containing an Si$-$C bond detected in IRC\,+10216 are SiC$_3$ \citep{app99}, SiC$_4$ \citep{ohi89}, SiCN \citep{Gue2000}, SiH$_3$CN, and CH$_3$SiH$_3$ \citep{agu2014,Cer2017}, but their abundances are low and the observed line profiles suggest that they are formed in the external layers of the envelope.

Most of our knowledge about the role of the SiC$_2$, SiC, and Si$_2$C molecules as gas-phase precursors of silicon carbide dust comes from the study of IRC\,+10216, but little is known about how widespread these molecules are in other carbon stars, what their relative abundances are, and what their role in the formation of SiC dust is. Here we present the results of a systematic survey carried out with the IRAM 30m telescope to observe these three molecules in a sample of 25 C-rich AGB stars. We report the detection of SiC$_2$ in most of the  sources and of the radical SiC in about half of them. The sample of stars and observational details are presented in Section~\ref{sec:observations} and the main results obtained from the observations in Section~\ref{sec:results}. In Section~\ref{sec:method} we describe the model and the excitation and radiative transfer calculations, and discuss the results from these calculations in Section~\ref{sec:results_model}. Finally, we discuss the implications of the derived SiC$_{2}$ abundances in Section~\ref{sec:sic2_abundances}, and present our conclusions in Section~\ref{sec:conclusions}.

\section{Observations} \label{sec:observations}

The observations were carried out in February and May 2016 with the IRAM 30m telescope, located at Pico Veleta (Spain). We selected a sample of 25 C-rich AGB stars with intense molecular emission, mainly based on the intensity of the HCN $J=\mbox{1-0}$ line \citep{lou1993,buj1994,sch2013}. The list of sources and their parameters are given in Table~\ref{table:sources}. The coordinates and LSR systemic velocities were taken from the literature \citep{olo1993,gro1996,gro2002,cer2000,san2012} and checked using the SIMBAD astronomical database\footnote{\tiny \texttt{http://simbad.u-strasbg.fr/Simbad}} \citep{wen2000}. For some sources, systemic velocities were not accurately known and thus we determined them from intense lines observed in this study (see below).

We used the E150 receiver in dual side band, with image rejections $>$10 dB, and observed the frequency ranges 138.5-146.3 GHz and 154.1-161.9 GHz (in the lower and upper side bands, respectively). This spectral setup was chosen to include several strong lines of SiC$_2$, various lines of Si$_2$C, and the $J=4-3$ rotational transition of the radical SiC in its $^3\Pi_2$ state. The strongest lines of these three species covered in the observed frequency range are listed in Table~\ref{table:lines}. In the IRC\,+10216 spectrum, the line intensity ratios of the three species are typically SiC$_2$:Si$_2$C:SiC $\sim$ 100:1:10 \citep{cer2000,cer2015}, and thus we expect that in the rest of the sources of the sample, SiC$_2$ will be the most easily detectable molecule, while Si$_2$C will display the weakest lines and thus will be the most difficult to detect. The adopted spectral range also covers  lines which are typically intense in carbon-rich circumstellar envelopes, such as HC$_3$N $J$ = 16-15 and $J$ = 17-16, C$_4$H $N$ = 15-14 and $N$ = 17-16, and SiS $J$ = 8-7.

The beam size of the telescope at these frequencies is in the range 15.0-17.5$''$. We used the wobbler-switching technique, with the secondary mirror nutating by 3$'$ at a rate of 0.5 Hz. The focus was regularly checked on Venus and Uranus and the pointing of the telescope was systematically checked on a nearby quasar before observing each AGB star. The error in the pointing is estimated to be 2-3$''$. The E150 receiver was connected to a fast Fourier transform spectrometer providing a spectral resolution of 200 kHz. The weather was good and stable during most of the observations, with typical amounts of precipitable water vapor of 1-3 mm and system temperatures ranging from 80 K to 150 K. The intensity scale, calibrated using two absorbers at different temperatures and the atmospheric transmission model (ATM) \citep{cer1985,par2001}, is expressed in terms of $T_A^*$, the antenna temperature corrected for atmospheric absorption and for antenna ohmic and spillover losses. To convert to main beam antenna temperature, $T_A^*$ has to be divided by 0.78 (the ratio of the main beam efficiency to the forward efficiency of the IRAM 30m telescope at the observed frequencies\footnote{\tiny \texttt{http://www.iram.es/IRAMES/mainWiki/Iram30mEfficiencies}}). The error in the intensities due to calibration is estimated to be $\sim$20 \%. 

The data were reduced using the software CLASS within the package GILDAS\footnote{\texttt{\tiny http://www.iram.fr/IRAMFR/GILDAS} \label{note:gildas}}. For each source, we averaged the spectra corresponding to the horizontal and vertical polarizations, subtracted a baseline consisting of a first-order polynomial, and smoothed the resulting spectrum to a spectral resolution of 1 MHz, which is good enough to spectrally resolve the lines whose widths in most sources are typically in the range 15-50 km s$^{-1}$ (7.5-25 MHz at the observed frequencies). Typical on-source integration times, after averaging horizontal and vertical polarizations, were $\sim$1 h for each source, resulting in $T_A^*$ rms noise levels per 1 MHz channel of 2-5 mK.

\begin{table}
\caption{Covered rotational transitions of SiC$_{2}$, Si$_2$C, and SiC}\label{table:lines}
\centering
\begin{tabular}{lccr}
\hline \hline
\multicolumn{1}{l}{Transition}  & \multicolumn{1}{c}{Frequency} & \multicolumn{1}{c}{$A_{ul}$} & \multicolumn{1}{c}{$E_u$} \\
\multicolumn{1}{l}{}                 & \multicolumn{1}{c}{(MHz)}  & \multicolumn{1}{c}{(s$^{-1}$)} & \multicolumn{1}{c}{(K)} \\
\hline
\multicolumn{4}{c}{SiC$_2$} \\
\hline
$6_{2,5}-5_{2,4}$ & 140920.171 & 7.65 $\times$ 10$^{-5}$ & 31.5 \\
$6_{4,3}-5_{4,2}$ & 141751.492 & 4.87 $\times$ 10$^{-5}$ & 55.0 \\
$6_{4,2}-5_{4,1}$ & 141755.360 & 4.87 $\times$ 10$^{-5}$ & 55.0 \\
$6_{2,4}-5_{2,3}$ & 145325.875 & 8.39 $\times$ 10$^{-5}$ & 32.0 \\
$7_{0,7}-6_{0,6}$ & 158499.228 & 1.23 $\times$ 10$^{-4}$ & 31.0 \\
\hline
\multicolumn{4}{c}{Si$_2$C} \\
\hline
$11_{1,11}-10_{0,10}$ & 144033.475 & 9.57 $\times$ 10$^{-6}$ & 29.3 \\
$22_{2,20}-22_{1,21}$ & 155600.100 & 1.54 $\times$ 10$^{-5}$ & 115.0 \\
$13_{1,13}-12_{0,12}$ & 157768.156 & 1.28 $\times$ 10$^{-5}$ & 39.3 \\
$20_{2,18}-20_{1,19}$ & 157959.754 & 1.54 $\times$ 10$^{-5}$ & 97.4 \\
$18_{2,16}-18_{1,17}$ & 160644.790 & 1.55 $\times$ 10$^{-5}$ & 81.4 \\
\hline
\multicolumn{4}{c}{SiC} \\
\hline
$^3\Pi_2$ $J=4-3$ & 157494.101 & 3.98 $\times$ 10$^{-5}$ & 13.2 \\
\hline
\end{tabular}
\end{table}

\section{Results from observations} \label{sec:results}

The observations resulted in the detection of SiC$_2$ in 22 out of the 25 targeted sources, i.e., all sources except UU\,Aur, R\,Lep, and ST\,Cam. The five rotational transitions of SiC$_2$ listed in Table~\ref{table:lines}, which have upper level energies in the range 31-55 K, were clearly detected in most of the 22 sources where SiC$_{2}$ was identified. The lines corresponding to the rotational transitions $6_{4,3}-5_{4,2}$ and $6_{4,2}-5_{4,1}$ appear blended together because line widths are larger than the frequency separation of the two transitions, although in most sources each of these lines could be fitted individually. The observed line profiles of SiC$_2$ are shown in Fig.~\ref{fig:sic2_lines} and the parameters derived from line fits using the \texttt{shell} method of CLASS$^{\ref{note:gildas}}$ are given in Table~\ref{table:sic2_lines}. We note that in U\,Cam, the line profiles of SiC$_2$ are consistent with emission arising exclusively from the present-day wind, rather than from the detached envelope (compare line profiles in Fig.~\ref{fig:sic2_lines} with those of CO in \citealt{sch2005}).

\begin{figure*}
\centering
\includegraphics[width=0.73\textwidth]{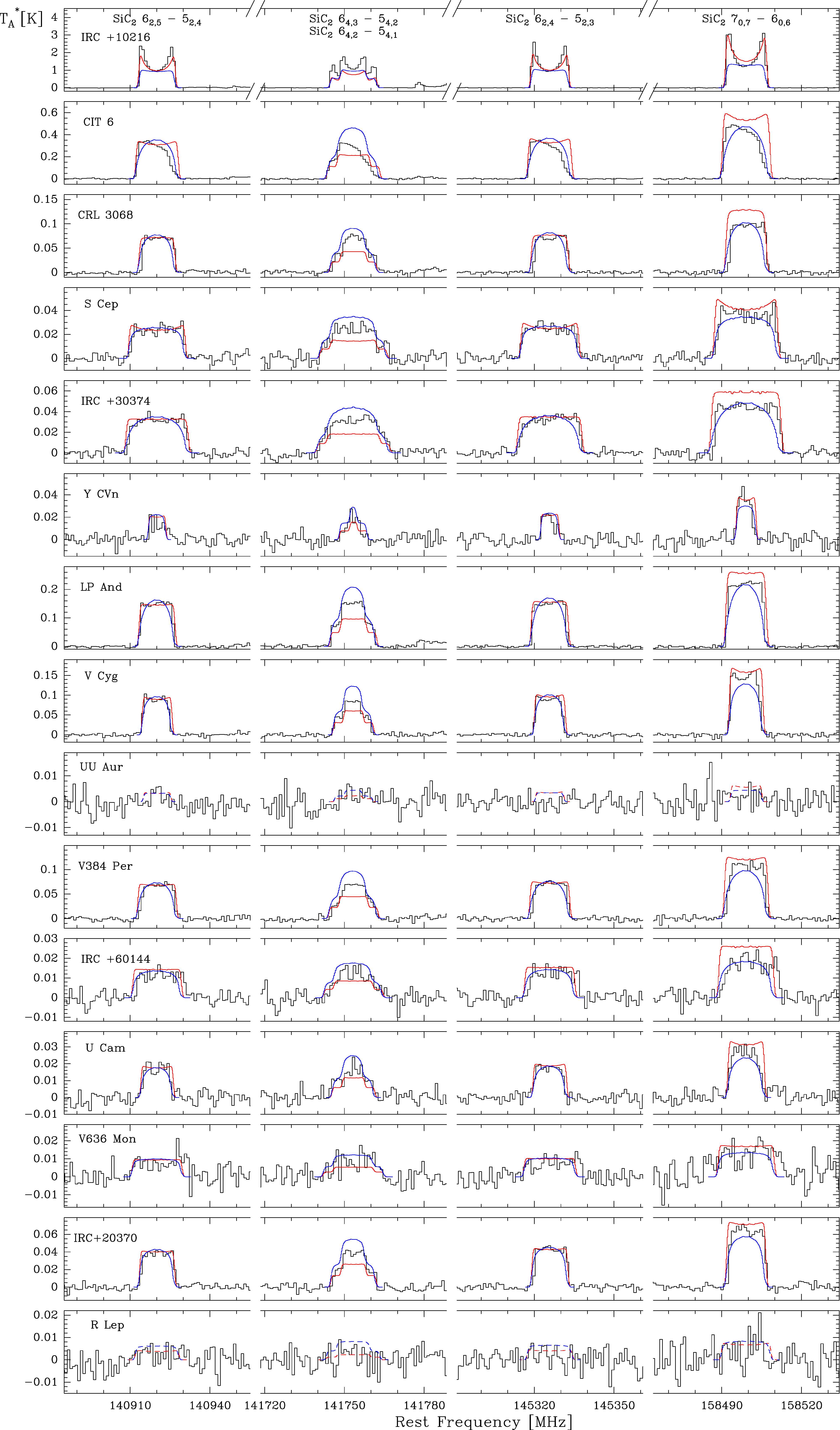}
\caption{Rotational lines of SiC$_{2}$ observed with the IRAM 30m telescope in the 25 target sources (black histograms). The spectral resolution of the observed lines is 1 MHz. The blue lines are the calculated emerging line profiles from the best-fit model using the LVG method. The red lines are the predicted line profiles assuming LTE excitation. Dashed lines correspond to calculated lines profiles for those sources where SiC$_{2}$ was not detected.}
\label{fig:sic2_lines}
\end{figure*}
\begin{figure*}
\ContinuedFloat
\captionsetup{list=off,format=cont}
\centering
\includegraphics[width=0.73\textwidth]{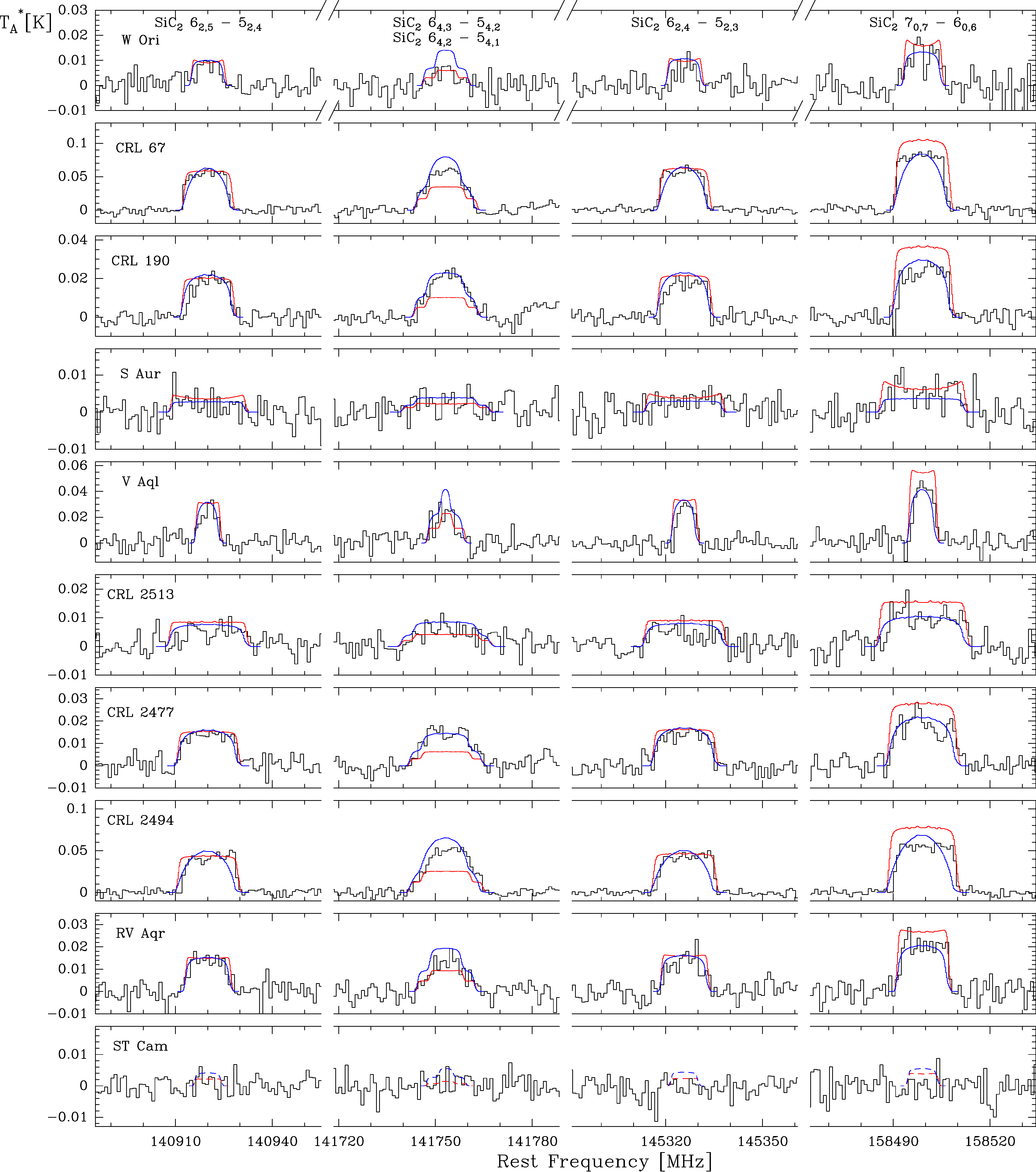}
\caption{}
\end{figure*}

\begin{figure} 
\centering
\includegraphics[width=0.98\columnwidth]{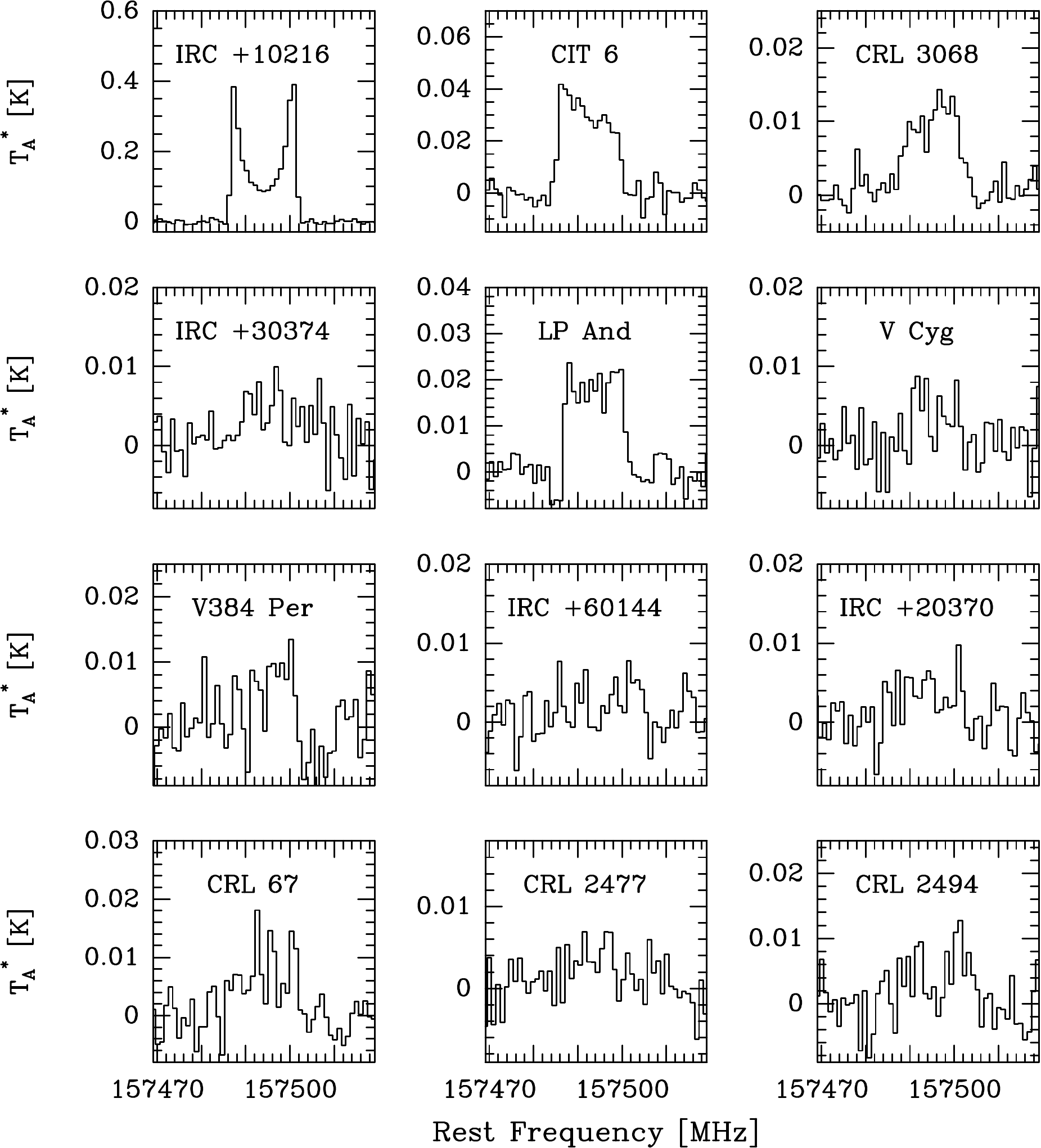}
\caption{SiC $^3\Pi_2$ $J=4-3$ line observed in 12 out of the 25 sources in our sample. The spectral resolution is 1 MHz.}
\label{fig:sic_emission_lines}
\end{figure}

\begin{figure}[ht]
\centering
\includegraphics[width=0.98\columnwidth]{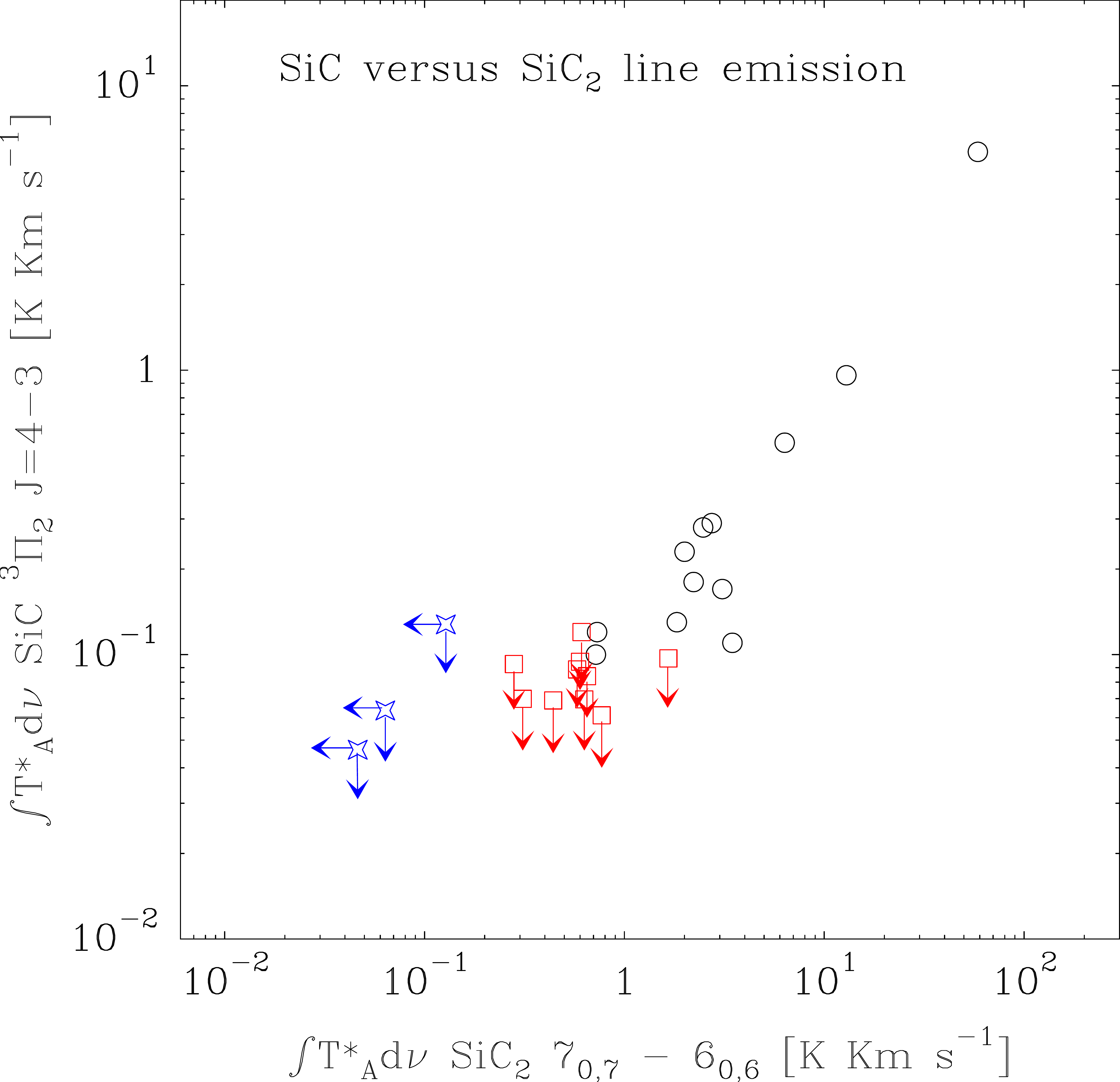}
\caption{Velocity-integrated intensity of the SiC $^3\Pi_2$ $J$ = 4-3 line versus the velocity-integrated intensity of the SiC$_{2}$ $7_{0,7}-6_{0,6}$ line. Black circles correspond to sources where both SiC$_2$ and SiC are detected, red squares to envelopes where SiC$_2$ is detected but not SiC, and blue stars to sources where neither SiC nor SiC$_2$ is detected.}
\label{fig:sic_vs_sic2}
\end{figure}

We also report the detection of SiC in 12 of the 25 targeted C-rich AGB envelopes (see Fig. \ref{fig:sic_emission_lines}). We note that prior to this study, this radical had been detected only in the C-star envelope IRC\,+10216 \citep{cer1989}. The SiC emission line is relatively strong in IRC\,+10216, CIT\,6, CRL\,3068, and LP\,And, while it is marginally detected in IRC\,+30374, V\,Cyg, V384\,Per, IRC\,+60144, IRC\,+20370, CRL\,67, CRL\,2477, and CRL\,2494. In Fig. \ref{fig:sic_vs_sic2} we compare the velocity-integrated intensity of the SiC $^3\Pi_2$ $J=4-3$ line with that of the SiC$_2$ $7_{0,7}-6_{0,6}$ line. We note that the SiC lines are brighter in the sources where SiC$_2$ is more intense, which suggests that the abundances of both species are correlated, as is expected if the radical SiC is a photodissociation product of SiC$_2$. This suggests that the nondetection of SiC in some of the sources is probably due to a lack of sensitivity and not due to a low fractional abundance of the radical. The related molecule Si$_2$C was not detected in any of the targeted sources, with the exception of IRC\,+10216, which at present is the only source in which this molecule has been identified \citep{cer2015}. In IRC\,+10216, the lines of Si$_2$C are typically 100 times less intense than those of SiC$_2$. This difference in intensity is not due to a significant difference in abundance, but it is related to the larger partition function of Si$_2$C and its lower dipole moment, as discussed in detail by \citet{cer2015}. If the same relative intensities between SiC$_2$ and Si$_2$C holds for the rest of C-star envelopes, the nondetection of disilicon carbide is very likely due to an insufficient sensitivity.

In addition to the lines of SiC$_2$ and SiC, we also detected within the frequency range covered 138.5-146.3 GHz and 154.1-161.9 GHz, emission lines of other species typically present in the spectra of carbon-rich AGB envelopes. For example, the $J=\mbox{16-15}$ and $J$ = 17-16 lines of HC$_3$N and the $J$ = 3-2 line of some isotopologues of carbon monosulfide, mainly C$^{34}$S and $^{13}$CS, are intense  and prevalent among most of the sources. The $J$ = 8-7 line of SiS is observed in many of the envelopes as one of the most intense lines, but it is not even detected in some of the other sources. The $N$ = 15-14 and $N$ = 17-16 lines of C$_4$H and the $N$ = 14-13 and $N$ = 16-15 of C$_3$N are also intense in some of the envelopes, although in many of them the lines of these two radicals are below the detection limit. In those sources where molecular emission is most intense, such as CIT\,6, CRL\,3068, and LP\,And, lines of other molecules such as $l$-C$_3$H, $c$-C$_3$H$_2$, C$_2$S, HC$_5$N, and rare isotopologues of CS, SiS, SiC$_2$, and HC$_3$N are detected. In a couple of sources, UU\,Aur and ST\,Cam, no molecular emission was observed above a detection threshold of a few mK in the covered frequency range.

The availability of strong lines of molecules such as HC$_3$N, SiS, C$_4$H, C$^{34}$S, and SiC$_2$ in our data allows us to derive accurate values of the velocity of the source $V_{\rm LSR}$ and of the terminal expansion velocity of the envelope $V_{\rm exp}$ in most of the sources. These two parameters are reported in the literature mainly from CO $J$ = 1-0 and $J$ = 2-1 lines \citep{lou1993, olo1993, gro1996, gro2002} with varying degrees of accuracy. We carried out a critical evaluation of the values of $V_{\rm LSR}$ and $V_{\rm exp}$ derived from our data and those in the literature. In cases where our lines have a well-defined shape, the values from our dataset were preferred, whereas when lines show a less clear shape, the values from literature were favored. The final values of $V_{\rm LSR}$ and $V_{\rm exp}$ adopted in this work are given in Table~\ref{table:sources}. 

In this article we focus on the analysis of SiC$_2$ and leave the interpretation of other molecules for future studies. Silicon dicarbide was detected in most of the targeted sources and we aim to model its emission and determine its abundance in each source to provide a global view of how abundant this molecule is in circumstellar envelopes around C-rich AGB stars.

\section{SiC$_2$ radiative transfer modeling} \label{sec:method}

\subsection{ Envelope model} \label{subsec:model}

We adopted a common physical scenario to model all sources consisting of a spherically symmetric envelope of gas and dust expanding with a constant velocity and mass-loss rate around a central AGB star. The parameters of the star and envelope adopted for the different sources are detailed in Table~\ref{table:sources}.

The central AGB star is characterized by an effective temperature $T_*$ and a luminosity $L_*$. The stellar radius $R_*$ is then determined by $T_*$ and $L_*$ using the Stefan--Boltzmann law. Effective temperatures for the photosphere of AGB stars, typically in the range 2000-3000 K, are difficult to estimate accurately. For most of the stars in our sample, we adopted the values of $T_*$ from studies where $T_*$ is derived by modeling the spectral energy distribution (SED) of each star (\citealt{sch2005, agu2012, sch2013, ram2014, dan2015}). In some cases, the adopted values of $T_*$ are taken from studies where the effective temperature was chosen to be a round number that fitted the SED reasonably well  (\citealt{gro1998, spe2009}). For those objects for which $T_*$ was not available from the literature, we assumed a typical value for AGB stars of 2500 K. Regarding the stellar luminosities $L_*$, we adopted values from the literature (see references in Table~\ref{table:sources}) in which they were mostly derived using the period-luminosity relation for Mira variables. 

The spherical envelope is described by the radial profile of various physical quantities, such as the gas density, the temperature of gas and dust, the expansion velocity, and the microturbulence velocity. The gas density as a function of the distance $r$ from the star is determined by the law of conservation of mass as
\begin{equation} \label{eq:density}
n_{g} = \frac{\dot{M}}{\overline{m}_{g}\, 4 \pi\, r^{2}\, V_e},
\end{equation}
where $n_{g}$ is the number of gas particles per unit volume, $\overline{m}_{g}$ is the average mass of gas particles, $\dot{M}$ is the mass-loss rate, and $V_e$ is the expansion velocity of the envelope. We assume that $V_e$ is equal to the terminal expansion velocity of the envelope $V_{\rm exp}$ across the entire envelope. This is certainly not true inside the acceleration region, but these inner layers contribute little to the emission of the SiC$_2$ lines observed here. We adopt $\mbox{$\overline{m}_{g}$= 2.3 amu}$, adequate for a gas composed mainly of H$_2$, He with a solar elemental abundance, and CO with an abundance of $\sim$10$^{-3}$ relative to H$_2$. Mass-loss rates were taken from the literature where $\dot{M}$ was determined by modeling observations of multiple CO lines (see references in Table \ref{table:sources}).

The radial structure of the gas kinetic temperature $T_k$ is described by a power law of the type
\begin{equation} \label{eq:tk}
T_k = T_{\star} \left(\frac{r}{R_{\star}}\right)^{-\delta},
\end{equation}
where $R_*$ is the stellar radius. In reality, the gas kinetic temperature is determined by a balance between the different heating and cooling processes at work, and cannot be accurately described by a single power law across the entire envelope. We do not aim to derive the gas kinetic temperature radial profile for each individual source, and thus for simplicity we adopt a uniform value of $\delta=0.7$, which is in line with findings from previous studies of circumstellar envelopes around AGB stars (e.g., \citealt{sch2001,deb2010,gue2017}).

The microturbulence velocity is assumed to be \SI{1}{\km\per\s} throughout the envelope, which is within the range of values  0.65 - \SI{1.5}{\km\per\s}, derived in the  literature for the C-rich envelope IRC\,+10216 \citep{ski1999,deb2012}; however, this parameter has  a limited impact when modeling the SiC$_2$ lines because it is well below the terminal expansion velocity of the wind for all sources.

We also consider the dusty component of the envelope, although, as will be discussed below, in our models dust plays a minor role in the excitation of the rotational lines of SiC$_2$ observed. We consider spherical grains of amorphous carbon with a radius of $\mbox{\SI{0.1}{\micro\metre}}$, a mass density of 2 g cm$^{-3}$, and optical properties from \citet{suh2000}. Dust grains are assumed to be present from the dust condensation radius $r_c$ with a constant gas-to-dust mass ratio $\Psi$. The radial structure of the dust temperature $T_d$ is assumed to be given by the power-law expression
\begin{equation} \label{eq:td}
T_d = T_d(r_c) \left(\frac{r}{r_c}\right)^{-\delta_d},
\end{equation}
where we assume $\delta_d=0.4$ based on theoretical expectations for carbonaceous grains (e.g., \citealt{hof2007}). This value is also close to that derived for IRC\,+10216 \citep{agu2012}. The values of $\Psi$, $r_c$, $T_d(r_c)$ were taken from the literature where they are typically derived by modeling the SED using photometric data, such as  IRAS and 2MASS fluxes and in some cases submillimeter data (\citealt{sch2006, sch2013, ram2014, dan2015}). In cases where those parameters were not available in the literature, we assumed a typical value of 5 R$_{\star}$ for $r_{c}$, 1200 K for $T_{\rm d}(r_{\rm c})$, and 300 for the gas-to-dust mass ratio $\Psi$. 

The adopted distances to the AGB stars were taken from the literature (see references in Table~\ref{table:sources}). The data include measurements from Hipparcos parallaxes and estimations based on bolometric magnitudes using the period-luminosity relation for Mira variables. When nothing else was available, the distances were estimated assuming $L_*$ = 10$^{4}$ L$_{\odot}$. 

\subsection{Excitation and radiative transfer calculations} \label{subsec:radiative_transfer}

We  performed  excitation and radiative transfer calculations to model the line emission of SiC$_{2}$ based on the multi-shell  large velocity gradient (LVG) method. The circumstellar envelope is divided into a number of concentric shells, each of which has  a characteristic set of physical properties and SiC$_{2}$ abundance, and statistical equilibrium equations are solved in each of them. In each shell, the contribution of the background radiation field (cosmic microwave background, stellar radiation, and thermal emission from surrounding dust) is included (see  \citealt{agu2012} for further details). For the current study, the LVG method provides a good compromise between the assumption of local thermal equilibrium (LTE) and more computationally expensive  nonlocal methods. Rate coefficients for the rotational excitation of SiC$_{2}$ through collisions have been computed theoretically \citep{cha2000}, which makes it worth using  more advanced methods than LTE. Our calculations indicate that the emission from the SiC$_2$ lines observed here arises from intermediate regions in the envelope where the rotational levels are not fully thermalized (see Section~\ref{sec:results_model}). Moreover, the observed lines of SiC$_2$ are optically thin in most sources, which implies that the LVG method should be accurate enough.

In the excitation calculations, we consider rotational levels up to $J=39$ and $K_a=20$ within the ground vibrational state of SiC$_2$ (i.e., a total number of 620 energy levels). Level energies and transition frequencies were calculated from the rotational constants reported by \citet{mul2012}, and line strengths for rotational transitions were computed from the dipole moment, 2.393$\pm$0.006 D, measured by \citet{sue1989}. The rate coefficients for excitation through inelastic collisions with H$_2$ and He were taken from \citet{cha2000}, who calculated rate coefficients for transitions between the first 40 rotational levels in the temperature range 25-125 K. At temperatures higher than 125 K, we adopted the theoretical rate coefficients calculated at 125 K and did not perform an extrapolation in temperature. This could introduce uncertainties in the excitation calculations. To evaluate the impact of this assumption, we carried out calculations in which we implemented a linear extrapolation in temperature of the rate coefficients at $\mbox{T$_{k}$ > 125 K}$ and verified that the calculated line profiles of the SiC$_2$ transitions observed were not sensitive to the particular choice of the collision rate coefficients at temperatures higher than 125 K. The reason is that, as is discussed in Section~\ref{sec:results_model}, most of the SiC$_2$ emission detected with the IRAM 30m telescope arise from intermediate regions of the envelope, where gas kinetic temperatures are in the range $\sim$ 50-300 K, i.e., not excessively far from the 25-125 K temperature range. Since the calculations of \citet{cha2000} only include the first 40 rotational levels, for transitions involving higher levels, the de-excitation rate coefficients were approximated using the  expression
\begin{equation}\label{eq:rate_coefficients}
\log{\gamma} = -10 - 0.25 \, (J'-J'') - 0.45 \, \vert K'_a-K''_a \vert,\end{equation}
where $\gamma$ is the rate coefficient in units of cm$^{3}$ s$^{-1}$, and $'$ and $''$ denote the upper and lower level, respectively. Equation~(\ref{eq:rate_coefficients}) provides a first-order approximation for asymmetric rotors, based on theoretical calculations carried out for the molecules SiC$_2$ \citep{cha2000}, SO$_2$ \citep{gre1995}, and HCO$_2^+$ \citep{ham2007}. 

\subsection{The SiC$_2$ radial abundance profile} \label{subsec:abundance_profile}

We consider that SiC$_2$ is formed close to the star with a given fractional abundance that remains constant throughout the envelope up to the outer regions of the envelope, where it is photodissociated by the ambient ultraviolet radiation field of the local interstellar medium. To compute the falloff of abundance due to photodissociation, we assume that the photodissociation rate of SiC$_2$ is given by the expression $\alpha\exp (-\beta A_V)$, where $A_V$ is the visual extinction. We adopt an  unattenuated rate of $\mbox{$\alpha=10^{-10}$ s$^{-1}$}$ and a dust shielding factor of $\beta=1.7$, values which are merely educated guesses \citep{mac1999} because the photodissociation cross section of SiC$_2$ is not well known. The radial variation of the SiC$_{2}$ abundance in the expanding envelope is then given by the differential equation \citep{jur1981,hug1982}
\begin{equation}\label{eq:abundance_equation}
\frac{df}{dr} = - \frac{\alpha}{V_{\rm exp}} \exp \left [-\left (\frac{r_d}{r} \right) \right] f,
\end{equation}
where $f$ is the fractional abundance of SiC$_2$ relative to H$_2$ and $r_{d}$ is a parameter that can be considered as a photodissociation radius and which is given by 
\begin{equation}
r_d = \frac{\beta\, \dot{M}}{4 \pi\, V_{\rm exp}\, \overline{m}_{g}\, 1.87\times 10^{21}},
\end{equation} 
where the numerical value is the canonical $N_{\rm H} / A_V$ ratio given by \citet{boh1978} for the local interstellar medium. The solution to Eq.~(\ref{eq:abundance_equation}) can be expressed as
\begin{equation} \label{eq:abundance_solution}
\begin{split}
\ln \frac{f}{f_{0}} = & - \frac{\alpha}{V_{\rm exp}} \Biggl\{ \left[ r_d\, E_i\, \left( -\frac{r_d}{r} \right) + r\,e^{(-r_{d}/r)} \right] \\
                             & - \left[ r_d\, E_{i}\, \left( -\frac{r_d}{R_*} \right) + R_*\, e^{(-r_d/R_*)} \right] \Biggl\},
\end{split}
\end{equation}
where $E_i$ is the exponential integral. This approach provides a simple and accurate way to take into account the abundance falloff due to photodissociation. Once the physical structure of the envelope is set and a given initial abundance $f_0$ is chosen, the abundance profile $f(r)$ becomes fully described by Eq.~(\ref{eq:abundance_solution}). We note that it is likely that SiC$_2$ can experience an abundance decline around the dust formation zone. The adopted abundance profile, however, does not include such a feature since we are tracing a specific region of the envelope, more specifically the intermediate one (see Section~\ref{sec:results_model}), and thus it is not possible to accurately derive the radial abundance profile from the very inner regions out to the outer envelope. In this work we thus determine the mean abundance of SiC$_2$ in the intermediate regions of the envelope of the studied sources.\\

In summary, to model the emission lines of SiC$_2$ and determine its abundance in the observed sources we constructed a model of the envelope for each source, as described in Section~\ref{subsec:model}, with the parameters given in Table~\ref{table:sources}. We then performed excitation and radiative transfer calculations, as explained in Section~\ref{subsec:radiative_transfer}, using the abundance profile described in Section~\ref{subsec:abundance_profile}. We varied the initial fractional abundance of SiC$_2$ relative to H$_2$, $f_0$ in Eq.~(\ref{eq:abundance_solution}), until the calculated line profiles matched the observed ones. We choose as the best-fit model the one that results in the best overall agreement between calculated and observed line profiles for the entire set of SiC$_2$ lines observed. In those cases where no lines of SiC$_2$ are detected, we derive upper limits to the abundance of SiC$_2$ by choosing the maximum abundance that results in line intensities compatible with the noise level of the observations.

\section{Results from SiC$_2$ radiative transfer modeling} \label{sec:results_model}

In most sources (see Fig.~\ref{fig:sic2_lines}), the shapes of the SiC$_2$ lines observed are nearly flat-topped, which is indicative of optically thin emission not resolved by the 15.0-17.5$''$ beam of the IRAM 30m telescope. One notable exception is IRC\,+10216, whose close proximity (130 pc) means that the emission is spatially resolved by the telescope beam, and the line profiles show a marked double-peaked character. The calculated line profiles resulting from our best-fit LVG model for each of the sources are shown in blue in Fig.~\ref{fig:sic2_lines}, where they are compared with the observed line profiles. The agreement between calculated and observed line shapes is good in most sources, except in IRC\,+10216, for which the LVG model produces a less marked U-shape than observed, and a few sources like LP\,And, for which the calculated lines are more curved than flat-topped; in other words, in some sources the LVG model seems to result in lines that are more optically thick than indicated by the observed line shapes. To investigate this, we  also ran models where we assumed LTE excitation for SiC$_2$. These models tend to produce line shapes that are in better agreement with the observed ones in sources like IRC\,+10216, where calculated line profiles have a  clearer U-shaped character, and LP\,And, where line shapes are more flat-topped (see red lines in Fig.~\ref{fig:sic2_lines}). This suggests that the LVG model lacks sufficient excitation for the levels involved in the observed SiC$_2$ transitions. There are various possible causes of the suspected lack of excitation, which we discuss below.

\begin{figure}
\centering
\includegraphics[width=\columnwidth]{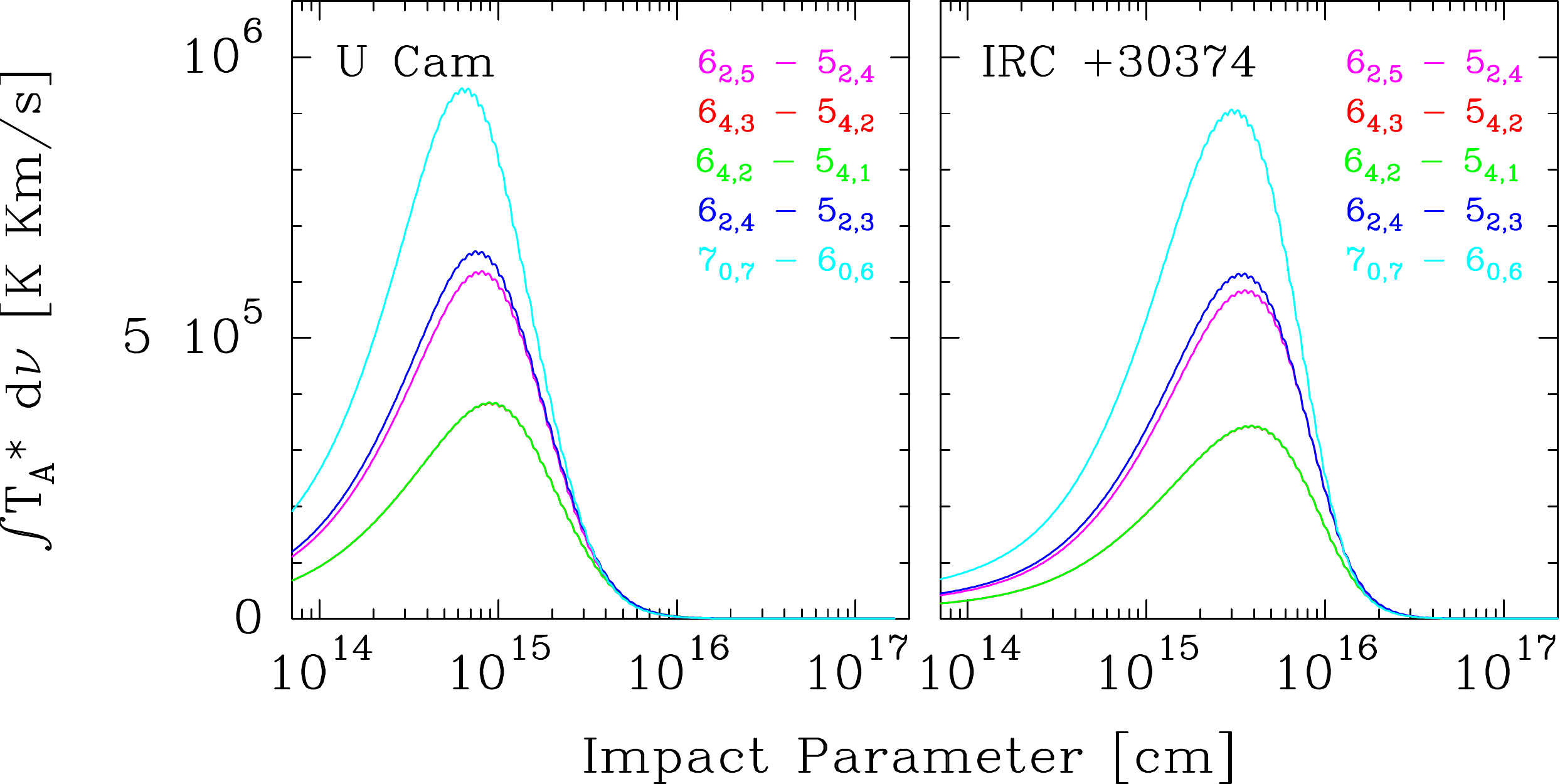}
\caption{Velocity-integrated intensity plotted as a function of impact parameter for the five observed SiC$_{2}$ lines in U\,Cam and IRC\,+30374. The transitions $6_{4,3}-5_{4,2}$ and $6_{4,2}-5_{4,1}$ have overlapping curves. For visualization reasons, the intensities of U\,Cam are multipled by a factor of 2.75.} \label{fig:impact_parameter}
\end{figure}

According to the LVG model, the excitation of the rotational levels of SiC$_2$ in the envelopes studied is dominated by inelastic collisions with H$_2$ and He. Even though thermal emission from dust is included in the model, it has little impact on the excitation of the observed SiC$_{2}$ rotational lines. The five lines of SiC$_2$ observed here involve upper levels with relatively low energies (31-55 K), and thus are preferentially excited in intermediate layers of the envelope where gas kinetic temperatures are of this order. In Fig.~\ref{fig:impact_parameter} we show the contribution to the velocity-integrated intensity of the five SiC$_2$ lines observed as a function of the impact parameter relative to the position of the star for two sources in our sample, U\,Cam and IRC\,+30374. These two envelopes are representative of very different mass-loss rates. While U\,Cam lies in the lower range, with $\dot{M}$ = $2\times10^{-7}$ M$_{\odot}$ yr$^{-1}$, IRC\,+30374 lies at the higher end, with $\dot{M}$ = $10^{-5}$ M$_{\odot}$ yr$^{-1}$. In U\,Cam, the maximum contribution to the line emission comes from regions at $\sim$10$^{15}$ cm, while in the case of IRC\,+30374 the regions around $\sim$ $10^{16}$ cm contribute the most to the observed emission; i.e.,  in envelopes with low mass-loss rates the SiC$_2$ rotational levels involved in the five observed transitions are only efficiently excited by collisions  in the inner regions (and thus most of the emission detected in these lines comes from such regions), while in envelopes with  high mass-loss rates the  densities are still high enough in the outer regions to excite the SiC$_2$ rotational levels and thus most emission comes from these outer regions. Among the studied sources, the LVG models indicate that emission from the five SiC$_2$ observed lines arises typically from intermediate regions of the envelope, at radial distances in the range 10$^{15}$-10$^{16}$ cm.

\begin{figure}
\centering
\includegraphics[width=\columnwidth]{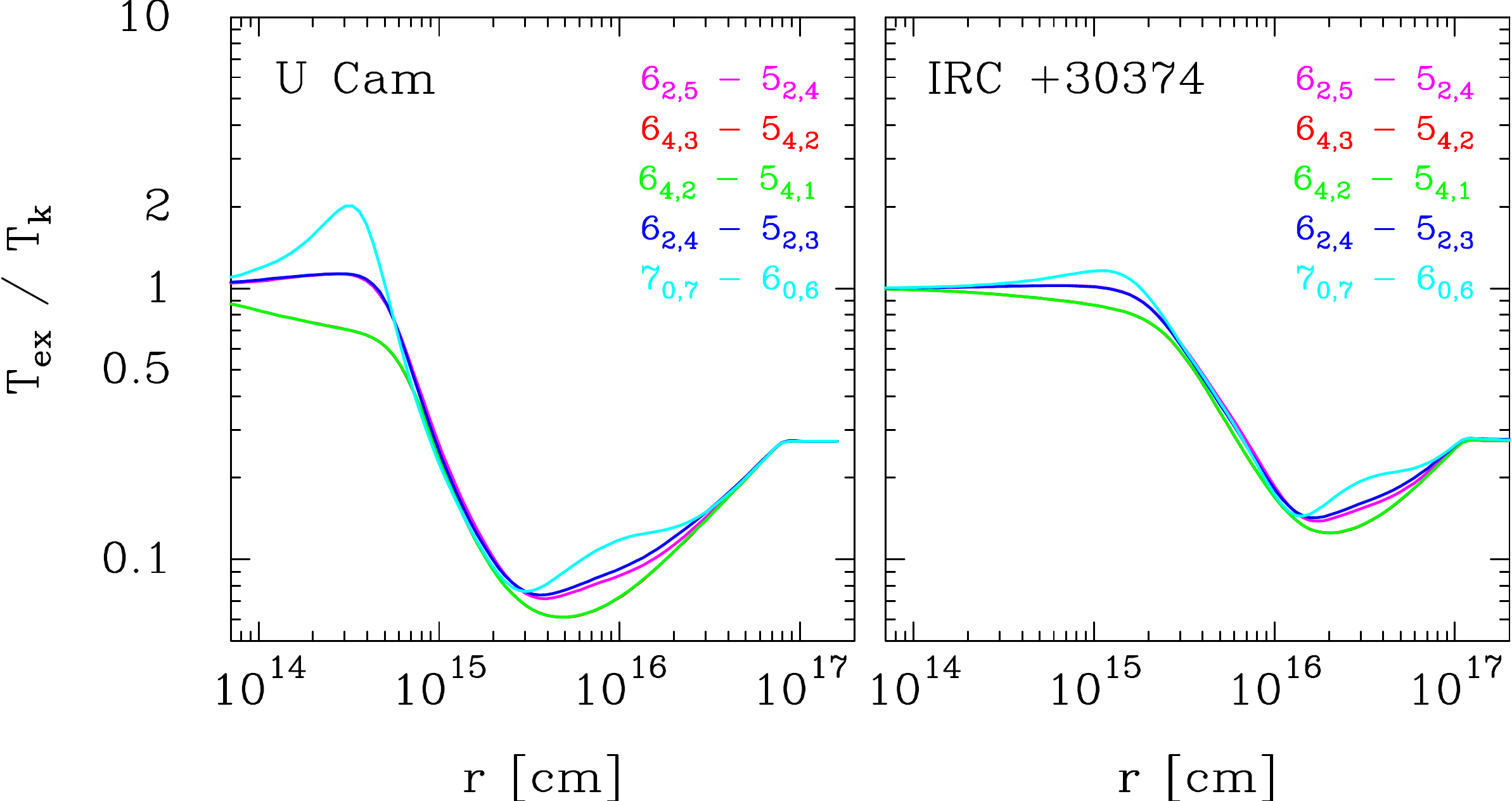}
\caption{Calculated ratio of excitation temperature to kinetic temperature ($T_{\rm ex}/T_{k}$) as a function of radius for the five observed rotational transitions of SiC$_{2}$ in U\,Cam and IRC\,+30374. The transitions $6_{4,3}-5_{4,2}$ and $6_{4,2}-5_{4,1}$ have overlapping curves.}
\label{fig:Texc_Tk}
\end{figure}

It is interesting to have a look at the excitation of SiC$_2$ predicted by the LVG model in the studied envelopes, in particular to the extent to which the rotational levels involved in the five observed transitions are close to or far from thermalization. In Fig. \ref{fig:Texc_Tk} we show the calculated ratio of excitation to gas kinetic temperature ($T_{\rm ex}/T_{k}$) for the five SiC$_2$ transitions as a function of radius for the envelopes U\,Cam and IRC\,+30374, which lie at low and high ranges, respectively, of mass-loss rates. We see that for most transitions $T_{\rm ex}/T_{k}$ = 1, i.e., rotational levels are thermalized, in the hot and dense inner regions, and that as the radius increases and the gas density decreases, the rotational levels become increasingly subthermally excited, as indicated by the fact that the ratio $T_{\rm ex}/T_{k}$ falls below unity. The much lower mass-loss rate of U\,Cam compared to IRC\,+30374 implies substantially lower densities in the envelope and thus in U\,Cam rotational populations start to deviate from thermalization at shorter radii than in IRC\,+30374. We note that in the intermediate region of the envelope from where most emission comes from, $\sim$10$^{15}$ cm in U\,Cam and $\sim$10$^{16}$ cm in IRC\,+30374 (see Fig.~\ref{fig:impact_parameter}), the SiC$_2$ rotational levels are mostly subthermally excited.

As commented above, for some of the sources, the LVG model could lack sufficient excitation for the levels involved in the observed transitions of SiC$_2$. One possible missing source of excitation could be related to the existence of shells with an enhanced density with respect to the surrounding media \citep{mau2000,cor2009,cern2015,agu2017,gue2017}. From models of IRC\,+10216 and LP\,And we notice that density enhancements above a factor of 10 are needed to start to reproduce adequately the observed line shapes. In IRC\,+10216, the shell--intershell density contrast is just 3 \citep{gue2017}. It is currently unknown whether episodic mass loss in other stars could be  abrupt enough to result in shell--intershell density contrasts above 10. We also note that uncertainties in the rate coefficients of SiC$_2$ excitation through inelastic collisions may also be at the origin of the suspected lack of excitation of SiC$_2$ in IRC\,+10216. In this sense, it would be interesting to revisit the collisional rate coefficients calculated by \cite{cha2000}. Another possible missing source of excitation in the LVG model could be infrared (IR) pumping, i.e., absorption of IR photons and pumping to excited vibrational states followed by spontaneous radiative decay to rotational levels in the ground vibrational state. This effect is not included in the model for SiC$_2$ (mainly because there is a complete lack of information about infrared intensities), although it is an important excitation mechanism of some molecules in IRC\,+10216 (e.g., \citealt{agu2006}) where the infrared flux is large \citep{cer1999}. However, vibrationally excited SiC$_2$ seems to be restricted to the inner layers, as indicated by the narrow lines from the vibrational states $\nu_3=1,2$ detected with ALMA \citep{cer2013}. Moreover, if IR pumping was playing an important role for SiC$_2$ we could expect to see the time variation of the line intensities found by \citet{cer2014} for other molecules, but the lines of SiC$_2$ show a rather constant intensity along the stellar period of CW\,Leo, which suggests that IR pumping is not important for this molecule \citep{cer2014}.

The differences between calculated and observed line shapes for some of the sources suggests that the LVG model may lack excitation for SiC$_2$, and that this could happen for other sources as well. We have thus run models assuming LTE excitation for all sources. The calculated line profiles are shown in red in Fig.~\ref{fig:sic2_lines}. Assuming LTE excitation throughout the whole envelope increases the excitation to outer radii, which leads to an increase in the emission size. Therefore, in general, the line opacities and fractional abundance of SiC$_2$ required to reproduce the observed intensities are lower when assuming LTE than when using the LVG method. Since it is unlikely that the rotational levels of SiC$_2$ are populated in LTE out to the outermost low-density regions of the envelope, the abundance of SiC$_2$ in the envelope of the observed sources is most likely between the values given by the LVG and the LTE models, which typically differ by a factor of $\sim$3 (see Table~\ref{table:abundances}).

\section{Discussion} \label{sec:sic2_abundances}

\begin{table}
\caption{Derived fractional abundances of SiC$_2$}\label{table:abundances}
\centering
\resizebox{\columnwidth}{!}{\begin{tabular}{lrcrr}
\hline \hline
\multicolumn{1}{l}{Name}  & \multicolumn{1}{c}{$\dot{M}$} & \multicolumn{1}{c}{$V_{\rm exp}$}  & \multicolumn{1}{c}{$f_0$(SiC$_2$)$_{\rm LVG}$} & \multicolumn{1}{c}{$f_0$(SiC$_2$)$_{\rm LTE}$}\\
\multicolumn{1}{c}{}  & \multicolumn{1}{c}{(M$_{\odot}$ yr$^{-1}$)} & \multicolumn{1}{c}{(km~s$^{-1}$)} & & \\
\hline
IRC\,+10216 & $2.0\times10^{-5}$   & 14.5  & $3.7\times10^{-7}$ &  $3.7\times10^{-7}$ \\
CIT\,6      & $6.0\times10^{-6}$   & 17    & $1.1\times10^{-5}$ &  $3.4\times10^{-6}$ \\
CRL\,3068   & $2.5\times10^{-5}$   & 14.5  & $1.9\times10^{-6}$ &  $7.2\times10^{-7}$\\ 
S\,Cep      & $1.2\times10^{-6}$   & 22.5  & $1.0\times10^{-5}$ &  $2.0\times10^{-6}$ \\
IRC\,+30374 & $1.0\times10^{-5}$   & 25    & $9.7\times10^{-6}$ &  $2.3\times10^{-6}$\\
Y\,CVn      & $1.5\times10^{-7}$   & 7     & $4.0\times10^{-6}$ &  $1.7\times10^{-6}$\\
LP\,And     & $7.0\times10^{-6}$   & 14.5  & $6.8\times10^{-6}$ &  $2.2\times10^{-6}$\\ 
V\,Cyg      & $1.6\times10^{-6}$   & 12    & $6.6\times10^{-6}$ &  $2.3\times10^{-6}$\\
UU\,Aur     & $2.4\times10^{-7}$   & 10.6  & $<1.0\times10^{-6}$&  $<3.5\times10^{-7}$\\ 
V384\,Per   & $2.3\times10^{-6}$   & 15.5  & $1.3\times10^{-5}$ &  $3.6\times10^{-6}$\\
IRC\,+60144 & $3.7\times10^{-6}$   & 19.5  & $6.0\times10^{-6}$ &  $1.8\times10^{-6}$\\
U\,Cam      & $2.0\times10^{-7}$   & 13    & $3.7\times10^{-5}$ &  $6.7\times10^{-6}$\\  
V636\,Mon   & $5.8\times10^{-6}$   & 20    & $1.7\times10^{-6}$ &  $5.1\times10^{-7}$\\
IRC\,+20370 & $3.0\times10^{-6}$   & 14    & $4.2\times10^{-6}$ &  $1.5\times10^{-6}$\\
R\,Lep      & $8.7\times10^{-7}$   & 17.5  & $<2.7\times10^{-6}$&  $<4.1\times10^{-7}$ \\  
W\,Ori      & $7.0\times10^{-8}$   & 11    & $1.4\times10^{-5}$ &  $2.4\times10^{-6}$\\
CRL\,67     & $1.1\times10^{-5}$   & 16    & $1.0\times10^{-5}$ &  $2.6\times10^{-6}$\\
CRL\,190    & $6.4\times10^{-5}$   & 17    & $8.8\times10^{-7}$ &  $2.9\times10^{-7}$\\
S\,Aur      & $4.0\times10^{-7}$   & 24.5  & $3.6\times10^{-6}$ &  $7.3\times10^{-7}$\\ 
V\,Aql      & $1.4\times10^{-7}$   & 8     & $2.0\times10^{-5}$ &  $7.3\times10^{-6}$\\
CRL\,2513   & $2.0\times10^{-5}$   & 25.5  & $1.6\times10^{-6}$ &  $5.3\times10^{-7}$\\
CRL\,2477   & $1.1\times10^{-4}$   & 20    & $6.0\times10^{-7}$ &  $1.9\times10^{-7}$\\
CRL\,2494   & $7.5\times10^{-6}$   & 20    & $2.7\times10^{-5}$ &  $4.9\times10^{-6}$\\
RV\,Aqr     & $2.3\times10^{-6}$   & 15    & $3.0\times10^{-6}$ &  $1.0\times10^{-6}$\\
ST\,Cam     & $1.3\times10^{-7}$   & 8.9   & $<4.0\times10^{-6}$&  $<6.0\times10^{-7}$ \\ 
\hline
\end{tabular}}
\end{table}

The fractional abundances of SiC$_2$ derived in the 25 studied envelopes are given in Table~\ref{table:abundances}. We give the abundances derived using the LVG method, while those obtained assuming LTE excitation are given in the last column.  The fractional abundances obtained using the LVG method range between $3.7\times10^{-7}$ and $3.7\times10^{-5}$ relative to H$_2$. Assuming that silicon has a solar elemental abundance in
AGB stars \citep{asp2009}, the maximum possible abundance of SiC$_2$ relative to H$_2$ is $6.5\times10^{-5}$. Therefore, the high fractional abundances of SiC$_2$ of a few times $10^{-5}$ derived in some of the envelopes imply that gaseous SiC$_2$ locks an important fraction of the available silicon, possibly making it a major reservoir of this element. We note, however, that if the abundances are closer to the values derived under LTE, the maximum abundances derived for SiC$_2$ are $\sim7\times10^{-6}$ relative to H$_2$, which would imply that SiC$_2$ locks at most $\sim10\%$ of the available silicon.

In most of the carbon-rich AGB envelopes in our sample, SiC$_2$ abundances are reported for the first time in this work. The only source where SiC$_2$ has been previously studied in detail is IRC\,+10216, which in fact has the lowest fractional abundance of SiC$_2$  among all the sources in our sample ($3.7\times10^{-7}$ relative to H$_2$) (see Table~\ref{table:abundances}). \cite{cer2010} observed 55 rotational transitions of SiC$_{2}$ in IRC\,+10216 using the HIFI spectrometer on board \textit{Herschel}. These authors derive an abundance of $2\times10^{-7}$ relative to H$_2$ in the inner regions of the envelope using an LTE radiative transfer analysis. Based on mm-wave interferometric observations, \cite{luc1995} derive an abundance relative to H$_2$ of $5\times10^{-7}$ for the inner regions, and more recently \citet{fon2014} inferred an abundance relative to H$_2$ of $8\times10^{-7}$ at the stellar surface decreasing down to $8\times10^{-8}$ at 20 R$_*$. The SiC$_2$ abundance derived in this work for IRC\,+10216 ($3.7\times10^{-7}$) is in good agreement with the values reported in these articles. 

\begin{figure*}
\centering
\includegraphics[width=0.95\columnwidth]{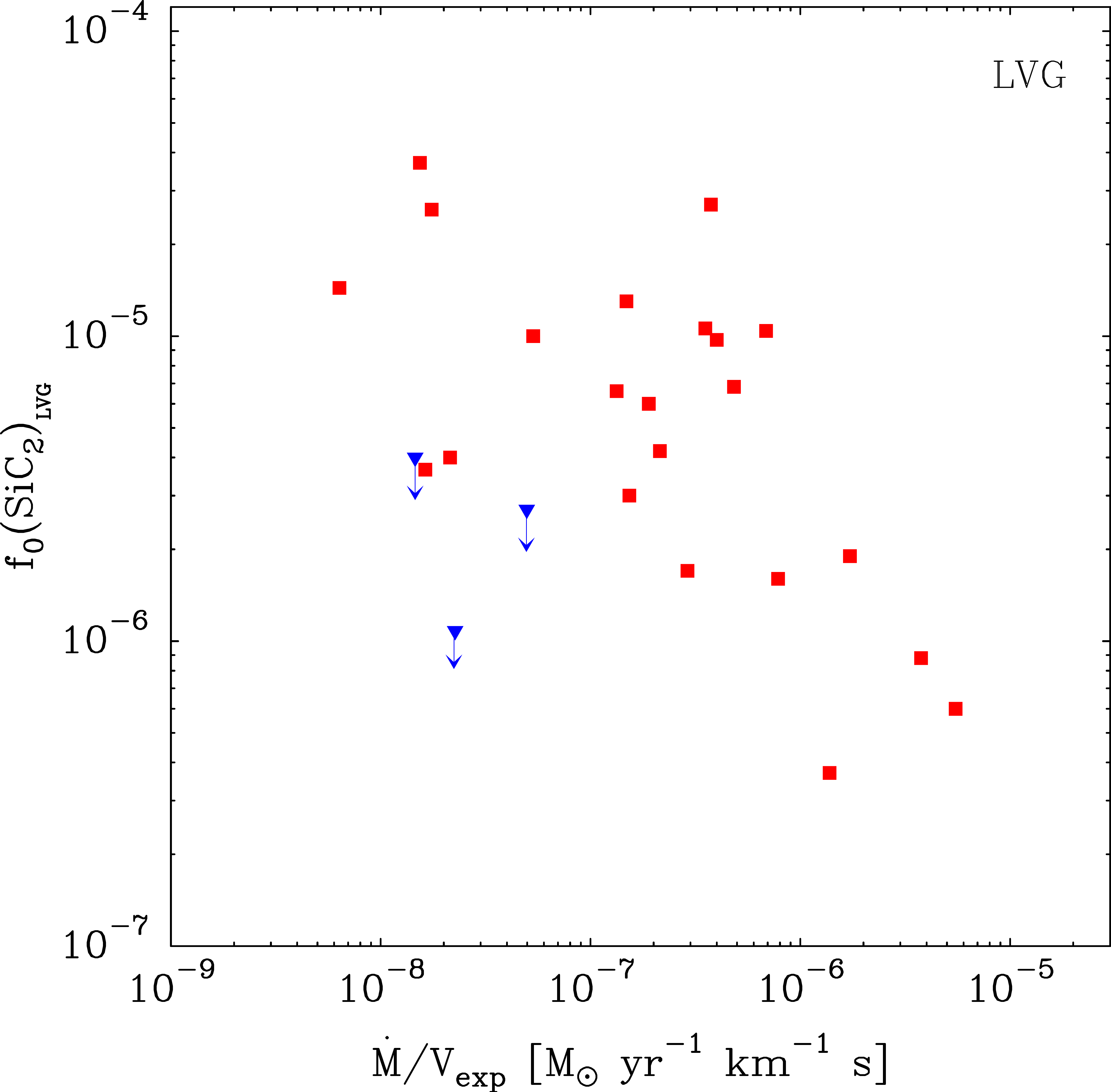} \hspace{1cm} \includegraphics[width=0.95\columnwidth]{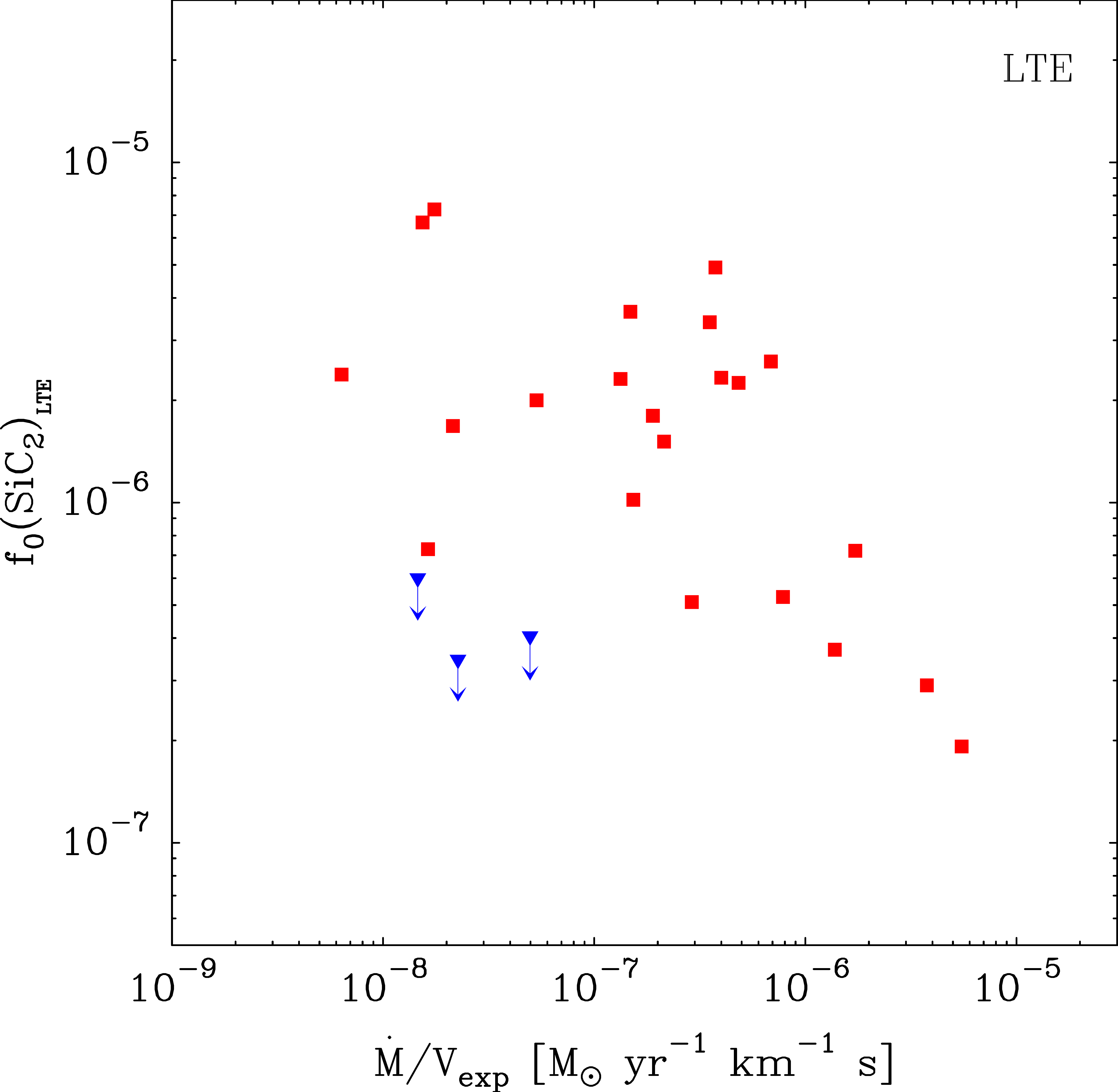}
\caption{Plot of the fractional abundances of SiC$_{2}$ ($f_{0}$) derived with the LVG method (left panel) and assuming LTE excitation (right panel) versus the envelope density proxy $\dot{M}/V_{\rm exp}$ for the 25 C-rich envelopes studied here. Blue downward triangles represent upper limits for $f_{0}$.}\label{fig:abun}
\end{figure*}

The fractional abundance of SiC$_2$ shows an interesting trend with either the mass-loss rate $\dot{M}$ or the density in the envelope, evaluated through the quantity $\dot{M}/V_{\rm exp}$. As shown in Fig.~\ref{fig:abun}, SiC$_2$ becomes less abundant as the density in the envelope increases. This trend is also present when the abundances derived assuming LTE are used. At this point, it is worth noting that the SiC$_2$ abundances derived in this work correspond to intermediate regions of the envelope. In a standard scenario describing the chemistry of an expanding envelope around an AGB star, the abundance with which a molecule is injected into the intermediate and outer envelope is set by the processes occurring in the inner regions. In this sense, the abundance of SiC$_2$ in the intermediate regions of the envelope is set by thermochemical equilibrium (TE) at the stellar surface and is possibly modified, i.e., driven out of thermochemical equilibrium, later on during the expansion by processes such as the formation of dust grains, shocks driven by the stellar pulsation, or even photochemical processes (e.g., \citealt{agu2010,che2012}). The observational finding of a decrease in the SiC$_2$ abundance with increasing gas density can therefore be interpreted in different ways.

\begin{figure}
\centering
\includegraphics[width=\columnwidth]{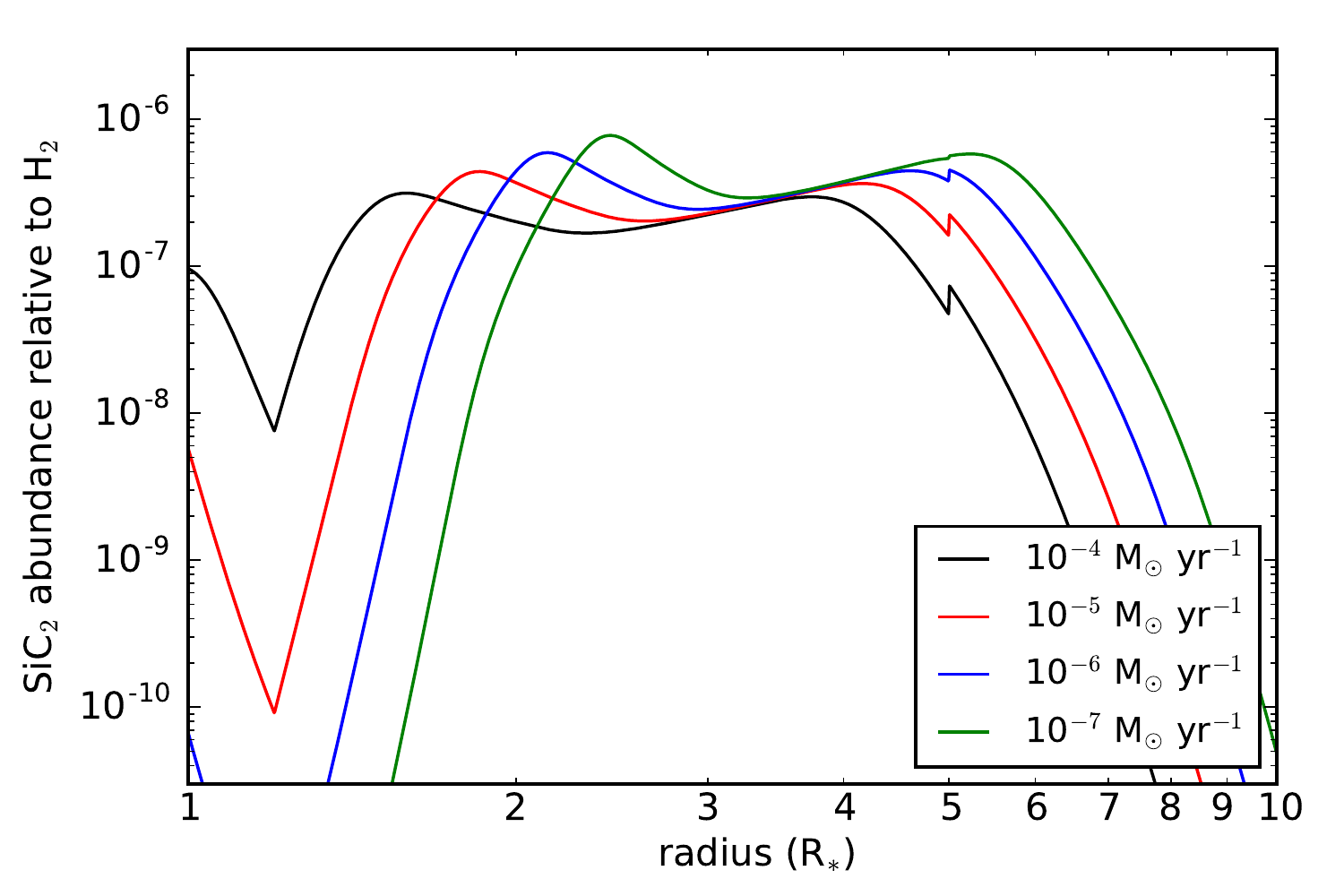}
\caption{Calculated fractional abundance of SiC$_{2}$ at thermochemical equilibrium as a function of distance to the star for various radial density profiles corresponding to mass-loss rates in the range 10$^{-7}$-10$^{-4}$ M$_{\odot}$ yr$^{-1}$.}\label{fig:abundance_lte}
\end{figure}

The observational trend could simply be a consequence of the way in which the TE abundance of SiC$_2$ depends on the density. To evaluate whether this could be a plausible explanation, we carried out thermochemical equilibrium calculations using the radial profiles of density and temperature of IRC\,+10216 (\citealt{agu2012}; see downward revision on the density profile by \citealt{cer2013}) and scaling  the density profile down or up depending on the mass-loss rate. In Fig.~\ref{fig:abundance_lte} we show the resulting TE abundance of SiC$_2$ within the first 10 R$_*$ around the star for mass-loss rates between 10$^{-7}$ and 10$^{-4}$ M$_{\odot}$ yr$^{-1}$. In the region where SiC$_2$ reaches its maximum TE abundance, between 2 and 5 R$_*$, the abundance of SiC$_2$ is not very sensitive to the density. Outside this region the abundance of SiC$_2$ shows a marked dependence on density. At radii $<2$ R$_*$, SiC$_2$ becomes more abundant with increasing density (contrary to the observational trend), while beyond 5 R$_*$ the SiC$_2$ abundance increases with decreasing density (as observed). At radii larger than 5 R$_*$, however, the TE abundance of SiC$_2$ experiences a drastic decline  to values well below those observed, and thus it is unlikely that such regions set the abundance that is injected into the expanding envelope. Although it is difficult to precisely locate the region where molecular abundances, in general, and that of SiC$_2$ in particular, quench to the TE value, such a region should be located around 2-3 R$_*$ \citep{agu2006,agu2012}, i.e., near the region where SiC$_2$ abundance is maximum and nearly insensitive to the density. We thus conclude that it is unlikely that the observational trend shown in Fig.~\ref{fig:abun} is caused by thermochemical equilibrium.

It  therefore seems that the observational finding of a decline in the SiC$_2$ abundance with increasing density is caused by some nonequilibrium process that takes place beyond the region where thermochemical equilibrium holds. The most natural explanation is that SiC$_2$ molecules deplete from the gas phase to incorporate into solid dust grains, a process that is favored at higher densities owing to the higher rate at which collisions between particles occur. We stress that since the SiC$_2$ abundances derived here correspond to intermediate regions of the envelope where dust formation has already taken place, they have to be considered as post-condensation abundances. Further support for this scenario comes from mm-wave interferometric observations of SiC$_2$ in the C-star envelope IRC\,+10216 \citep{luc1995,fon2014,vel2015}, which shows that SiC$_2$ is present in regions close to the star, then experiences a marked abundance decline at 10-20 R$_*$ (very likely due to condensation onto dust grains), and appears again in the outer envelope (probably as a result of the interaction between the UV radiation field and the envelope).

\begin{figure}
\centering
\includegraphics[width=0.97\columnwidth]{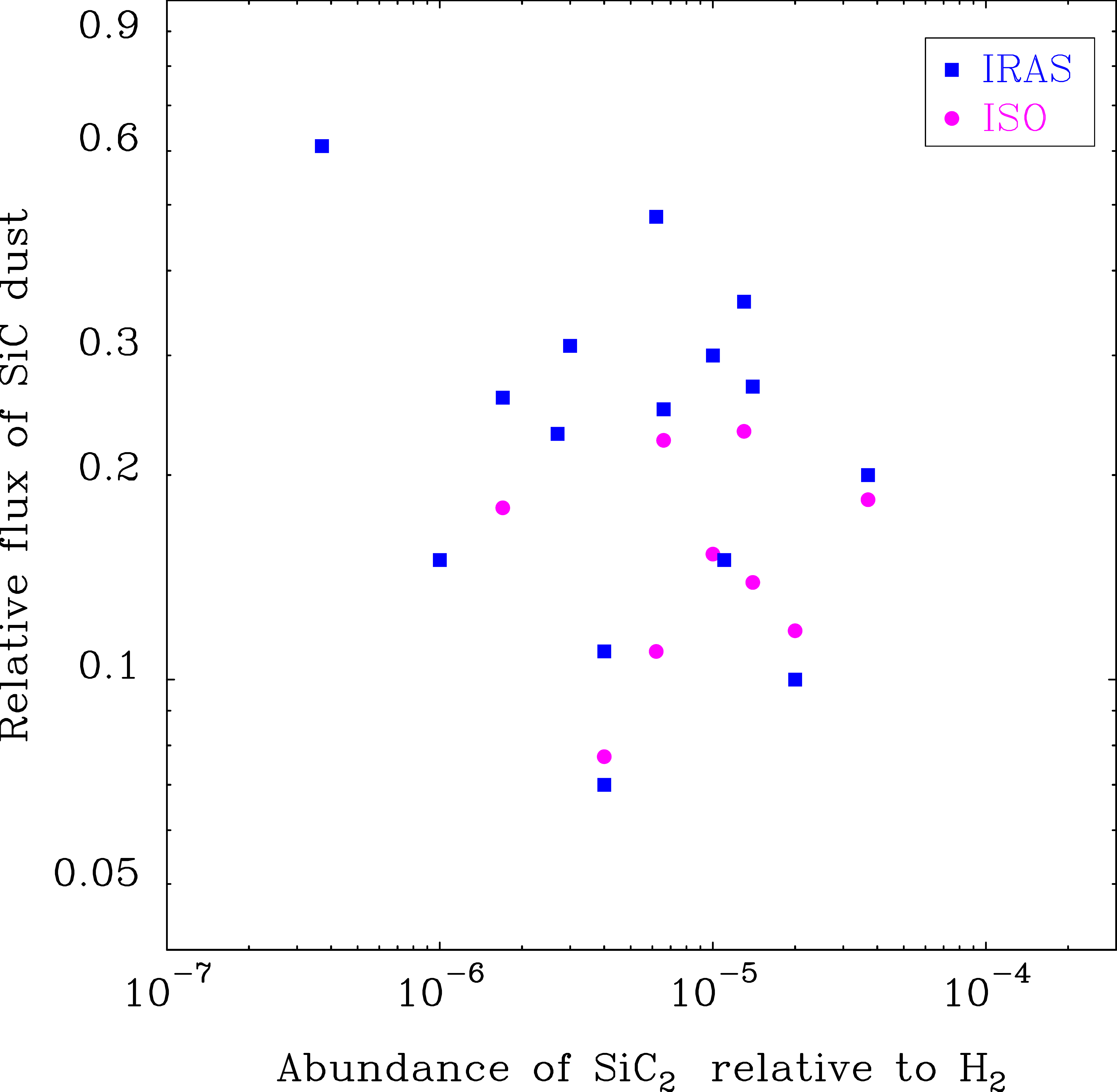}
\caption{Relative integrated flux of the SiC dust feature at \SI{11.3}{\micro\metre} taken from the literature versus the fractional abundance of SiC$_2$ derived in this work.}\label{fig:dustfeature}
\end{figure}

In order to further evaluate the hypothesis that the decline in the abundance of SiC$_2$ with increasing envelope density is caused by a more efficient incorporation of SiC$_2$ on silicon carbide dust, we collected information on infrared IRAS and ISO data for the sources in our sample that exhibit the SiC dust emission feature at \SI{11.3}{\micro\metre}. Among the 25 sources in our sample, for 15 of them \citet{slo98} have analyzed the IRAS LRS spectra and for 9 of them \citet{yan2004} have studied the ISO SWS spectra. These authors determine the relative flux of SiC dust as the ratio of the integrated flux of the \SI{11.3}{\micro\metre} emission feature (after continuum substraction) divided by the integrated flux of the continuum\footnote{The main difference between the two studies lies in the integration range, which is 7.67-\SI{14.03}{\micro\metre} for the IRAS spectra and 9-\SI{13.6}{\micro\metre} for the ISO data. For the sources that were observed with both telescopes, the main causes of the differences between the IRAS and ISO relative fluxes of SiC dust are the different integration ranges and the specific error in the spectra taken by each telescope.}. In Fig.~\ref{fig:dustfeature} we plot the relative integrated flux of SiC dust versus the fractional abundance of SiC$_2$ for the sources in our sample which have IRAS or ISO data. If the relative flux of SiC dust is a good proxy of the amount of silicon carbide dust, and if the hypothesis that SiC$_2$ is a gas-phase precursor of SiC dust is correct, one would expect to see a trend where the relative flux of SiC dust increases as the gas-phase abundance of SiC$_2$ decreases. This trend is not visible in Fig.~\ref{fig:dustfeature}. We note, however, that the relative flux of the \SI{11.3}{\micro\metre} SiC band is an observable quantity that may not necessarily be a good proxy of the mass of silicon carbide dust in the envelope, the derivation of which requires a detailed radiative transfer analysis that includes a thorough description of the chemical composition and temperature of dust throughout the envelope. At this point, it therefore remains inconclusive whether or not there is a clear observational relation between the abundances of SiC$_2$ gas and SiC dust in C-rich envelopes. We  plan to address this issue in a future study.

The  question of identifying the main gas-phase precursors of dust grains in the ejecta of AGB stars has been, and continues to be, an exciting scientific topic. The problem has been addressed from different perspectives, mainly in the context of oxygen-rich AGB stars. For example, in a series of studies the abundance of SiO has been investigated in detail in a wide sample of envelopes around M stars \citep{gon2003}, C stars \citep{sch2006}, and stars of S type \citep{ram2009}. Results from these studies show a clear trend of SiO becoming less abundant as the density of the envelope increases, a similar behavior to that found in this work for SiC$_2$ in C-rich envelopes. The observed trend for SiO has been also interpreted in terms of a more efficient adsorption of SiO onto dust grains for high envelope densities. It  has been noted, however,  that in the case of S stars, the trend is not as clear as in M and C stars, although this may be due to the low number of high mass-loss S stars observed  (see \citealt{ram2009}). A similar study carried out on the abundance of SiS in envelopes around O- and C-rich stars \citep{sch2007} did not find a clear correlation between the fractional abundance of SiS and the density of the envelope, suggesting that this molecule is not affected as much as SiO by freeze-out onto dust grains. The conclusions regarding SiS,  however, are limited by the low number of O- and C-rich stars studied.

In a recent study, \citet{liu2017} investigated the relation between the intensities of SiO maser and silicate dust emission in a sample of O-rich AGB and post-AGB stars in an attempt to relate the abundance of SiO, a highly plausible gas-phase precursor of silicates, with the amount of silicate dust. These authors find a positive correlation between the velocity-integrated intensity of SiO maser and the emission power of silicate dust, which is not against the hypothesis that SiO is a gas-phase precursor of silicate dust. It is important to note that in envelopes around O-rich AGB stars, SiO masers probe inner regions to where dust forms \citep{reid97,dan94}, and thus the more abundant the precursor SiO  in the maser formation region is, the higher the amount of silicate dust can later  be formed. In any case, given the peculiar nature of maser emission (e.g., \citealt{gra2009}) and the fact that the observables are not a straightforward proxy of abundances, it is not clear whether the observed trend can be interpreted in terms of a higher fractional abundance of SiO in envelopes with a higher amount of silicate dust.

Various studies carried out in recent years have used ALMA to investigate the potential role of titanium and aluminum oxides as gas-phase precursors of inorganic dust in oxygen-rich evolved stars. In M stars, TiO$_2$ and Al$_2$O$_3$ are predicted to be among the first solid compounds to condense out of the gas phase \citep{gai1998} and thus are ideal candidates to initiate the dust formation process. \citet{deb2015} and more recently \citet{kam2017} have observed TiO and TiO$_2$ in the surroundings of the evolved stars VY Canis Majoris and Mira and found that these molecules extend out to several stellar radii from the star, and thus are unlikely to act as gas-phase precursors of TiO$_2$ dust. In a similar study, \citet{kam2016} observed AlO around the star Mira, although they could not quantitatively assess the role of AlO as a gas-phase precursor of Al$_2$O$_3$ dust. ALMA is indeed a very promising tool for  identifying gas-phase precursors of dust in evolved stars, but  results are still not conclusive.

\section{Conclusions} \label{sec:conclusions}

In this work we used the IRAM 30m telescope to survey a sample of 25 C-rich circumstellar envelopes and search for rotational emission from the molecules SiC$_2$, SiC, and Si$_2$C. We detected SiC$_2$ in most of the sources and SiC in about half of them, while Si$_2$C was not detected in any of the sources with the exception of IRC\,+10216. We carried out excitation and radiative transfer calculations to derive SiC$_2$ fractional abundances. We found a clear trend in which SiC$_2$ becomes less abundant as the envelope density increases. We interpret this result as  evidence of efficient incorporation of SiC$_2$ onto dust grains, a process that is more efficient at high densities because collisions between particles and coagulation processes become faster. The ring molecule SiC$_2$ thus emerges as a very likely gas-phase precursor in the process of formation of SiC dust in envelopes around C-rich AGB stars; therefore,  SiC$_2$  seems to behave similarly to SiO, which has been found to deplete from the gas phase as the density in the envelope increases. The search for the gas-phase building blocks of dust grains in both carbon- and oxygen-rich AGB stars is a challenging scientific objective, in which astronomical observations still have to provide many answers. Observing other molecules containing refractory elements is necessary in order to understand which of the gas-phase molecules that are present in the atmospheres of AGB stars serve as precursor material to feed the process of dust grain formation.

\begin{acknowledgements}

We thank the IRAM 30m staff for their help during the observations. This research has made use of the SIMBAD database, operated at CDS, Strasbourg, France. We acknowledge funding support from the European Research Council (ERC Grant 610256: NANOCOSMOS) and from Spanish MINECO through grants AYA2012-32032 and AYA2016-75066-C2-1-P. M.A. thanks Spanish MINECO for funding support through the Ram\'on y Cajal programme (RyC-2014-16277).

\end{acknowledgements}

\nocite{Gua2006}
\nocite{Gua2013}
\bibliographystyle{aa} 
\bibliography{mybib} 

\setcounter{table}{0}
\renewcommand{\thetable}{A\arabic{table}}
\begin{appendices}
\section{}
\longtab{
\begin{longtable}{lcccc}
\caption{Observed line parameters of SiC$_2$} \label{table:sic2_lines} \\\hline \hline
\multicolumn{1}{l}{Line} & \multicolumn{1}{c}{$\nu_{calc}$} & \multicolumn{1}{c}{$\nu_{obs}$} & \multicolumn{1}{c}{$V_{e}$}      & \multicolumn{1}{c}{$\int T_A^* dv$} \\
& \multicolumn{1}{c}{(MHz)}           & \multicolumn{1}{c}{(MHz)}          & \multicolumn{1}{c}{(km s$^{-1}$)} & \multicolumn{1}{c}{(K km s$^{-1}$)} \\
\hline
\endfirsthead
\caption{Continued.} \\
\hline
\multicolumn{1}{l}{Line} & \multicolumn{1}{c}{$\nu_{calc}$} & \multicolumn{1}{c}{$\nu_{obs}$} & \multicolumn{1}{c}{$V_{e}$}  & \multicolumn{1}{c}{$\int T_A^* dv$} \\
& \multicolumn{1}{c}{(MHz)}  & \multicolumn{1}{c}{(MHz)}  & \multicolumn{1}{c}{(km s$^{-1}$)} & \multicolumn{1}{c}{(K km s$^{-1}$)} \\
\hline
\endhead
\hline
\endfoot
\multicolumn{5}{c}{IRC\,+10216} \\
\hline
 $6_{2,5}-5_{2,4}$        & 140920.171  & 140920.1(1)     & 13.8(1)       & 41.2(4) \\
 $6_{4,3}-5_{4,2}$        & 141751.492  & 141751.3(1)     & 13.8(1)       & 21.5(2)$^a$ \\
 $6_{4,2}-5_{4,1}$        & 141755.360  & 141755.2(1)     & 13.8(1)       & 22.2(2)$^a$\\
 $6_{2,4}-5_{2,3}$        & 145325.875  & 145325.8(1)     & 14.3(1)       & 42.4(4) \\
 $7_{0,7}-6_{0,6}$        & 158499.228  & 158499.1(1)     & 15.3(1)       & 58.7(5) \\
\hline
\multicolumn{5}{c}{CIT\,6} \\
\hline
 $6_{2,5}-5_{2,4}$        & 140920.171  & 140920.5(10)     & 16.2(8)         & 8.2(8)\\
 $6_{4,3}-5_{4,2}$        & 141751.492  & 141751.1(2)      & 15.4(8)         & 5.4(5)$^a$ \\
 $6_{4,2}-5_{4,1}$        & 141755.360  & 141754.9(3)      & 14.0(8)         & 2.7(3)$^a$\\
 $6_{2,4}-5_{2,3}$        & 145325.875  & 145325.0(10)     & 13.8(10)        & 8.0(8) \\
 $7_{0,7}-6_{0,6}$        & 158499.228  & 158498.5(10)     & 15.1(8)         & 12.9(13) \\
\hline
\multicolumn{5}{c}{CRL\,3068} \\
\hline
 $6_{2,5}-5_{2,4}$        & 140920.171  & 140920.5(1)   & 12.8(8)    & 1.78(2) \\
 $6_{4,3}-5_{4,2}$        & 141751.492  & 141751.3(1)   & 12.8(4)    & 1.07(1)$^a$ \\
 $6_{4,2}-5_{4,1}$        & 141755.360  & 141755.3(1)   & 12.8(4)    & 1.02(1)$^a$\\
 $6_{2,4}-5_{2,3}$        & 145325.875  & 145325.7(1)   & 13.2(6)    & 1.84(2) \\
 $7_{0,7}-6_{0,6}$        & 158499.228  & 158499.2(2)   & 14.5(8)    & 2.73(3) \\
\hline
\multicolumn{5}{c}{S\,Cep} \\
\hline
 $6_{2,5}-5_{2,4}$        & 140920.171  & 140920.3(5)         & 20.5(4)       & 0.94(9) \\
 $6_{4,3}-5_{4,2}$        & 141751.492  & 141751.3(5)         & 21.8(1)       & 0.63(6)$^a$\\
 $6_{4,2}-5_{4,1}$        & 141755.360  & 141755.9(5)         & 22.4(1)       & 0.49(5)$^a$\\
 $6_{2,4}-5_{2,3}$        & 145325.875  & 145325.5(5)         & 21.7(8)       & 1.06(10) \\
 $7_{0,7}-6_{0,6}$        & 158499.228  & 158499.9(2)         & 23.0(5)       & 1.66(16) \\
\hline
\multicolumn{5}{c}{IRC\,+30374} \\
\hline
 $6_{2,5}-5_{2,4}$        & 140920.171  & 140920.4(1)        & 23.0(8)     & 1.43(14) \\
 $6_{4,3}-5_{4,2}$        & 141751.492  & 141751.8(3)        & 22.8(4)     & 0.84(8)$^a$\\
 $6_{4,2}-5_{4,1}$        & 141755.360  & 141755.5(3)        & 22.9(4)     & 0.62(6)$^a$\\
 $6_{2,4}-5_{2,3}$        & 145325.875  & 145325.9(1)        & 23.4(4)     & 1.50(4) \\
 $7_{0,7}-6_{0,6}$        & 158499.228  & 158499.4(1)        & 25.5(8)     & 2.22(2) \\
\hline
\multicolumn{5}{c}{Y\,CVn} \\
\hline
 $6_{2,5}-5_{2,4}$        & 140920.171  &  140921.7(10)      & 8.6(10)     & 0.23(4) \\
 $6_{4,3}-5_{4,2}$        & 141751.492  &  141751.7(2)       & 7.3(8)      & 0.16(2)$^a$ \\
 $6_{4,2}-5_{4,1}$        & 141755.360  &  141755.7(2)       & 7.5(6)      & 0.12(2)$^a$ \\
 $6_{2,4}-5_{2,3}$        & 145325.875  &  145326.3(10)      & 7.1(10)     & 0.26(5)  \\
 $7_{0,7}-6_{0,6}$        & 158499.228  &  158499.1(5)       & 6.6(5)      & 0.44(9) \\
\hline
\multicolumn{5}{c}{LP\,And} \\
\hline
 $6_{2,5}-5_{2,4}$        & 140920.171  & 140920.3(1)       & 13.0(8)     & 3.86(4) \\
 $6_{4,3}-5_{4,2}$        & 141751.492  & 141751.6(2)       & 13.1(8)     & 1.93(2)$^a$ \\
 $6_{4,2}-5_{4,1}$        & 141755.360  & 141755.4(2)       & 13.2(8)     & 2.06(2)$^a$\\
 $6_{2,4}-5_{2,3}$        & 145325.875  & 145325.9(1)       & 13.4(8)     & 3.99(4) \\
 $7_{0,7}-6_{0,6}$        & 158499.228  & 158499.3(1)       & 14.6(8)     & 6.34(6) \\
\hline
\multicolumn{5}{c}{V\,Cyg} \\
\hline
 $6_{2,5}-5_{2,4}$          & 140920.171  & 140920.3(1)      & 10.4(8)    & 1.87(2) \\
 $6_{4,3}-5_{4,2}$          & 141751.492  & 141751.7(2)      & 10.8(4)    & 0.91(2)$^a$ \\
 $6_{4,2}-5_{4,1}$          & 141755.360  & 141755.4(2)      & 10.6(4)    & 0.91(2)$^a$\\
 $6_{2,4}-5_{2,3}$          & 145325.875  & 145326.1(3)      & 11.0(8)    & 1.95(2) \\
 $7_{0,7}-6_{0,6}$          & 158499.228  & 158499.4(1)      & 11.8(8)    & 3.47(3) \\
\hline
\multicolumn{5}{c}{V384\,Per} \\
\hline
 $6_{2,5}-5_{2,4}$        & 140920.171  & 140920.2(1)      & 13.5(4)        & 1.80(2) \\
 $6_{4,3}-5_{4,2}$        & 141751.492  & 141751.4(1)      & 13.9(2)        & 0.80(1)$^a$ \\
 $6_{4,2}-5_{4,1}$        & 141755.360  & 141755.1(1)      & 13.8(2)        & 1.04(2)$^a$ \\
 $6_{2,4}-5_{2,3}$        & 145325.875  & 145326.0(1)      & 13.6(8)        & 1.88(2) \\
 $7_{0,7}-6_{0,6}$        & 158499.228  & 158499.3(1)      & 15.1(1)        & 3.09(3) \\
\hline
\multicolumn{5}{c}{IRC\,+60144} \\
\hline
 $6_{2,5}-5_{2,4}$        & 140920.171  & 140920.4(5)      & 19.1(8)   & 0.55(5) \\
 $6_{4,3}-5_{4,2}$        & 141751.492  & 141751.3(1)      & 18.5(2)   & 0.37(4)$^a$  \\
 $6_{4,2}-5_{4,1}$        & 141755.360  & 141755.4(4)      & 19.9(3)   & 0.25(2)$^a$  \\
 $6_{2,4}-5_{2,3}$        & 145325.875  & 145325.9(2)      & 19.7(1)   & 0.25(2) \\
 $7_{0,7}-6_{0,6}$        & 158499.228  & 158499.4(2)      & 21.5(4)   & 0.73(7) \\
\hline
\multicolumn{5}{c}{U\,Cam} \\
\hline
 $6_{2,5}-5_{2,4}$        & 140920.171  & 140920.8(5)         & 11.2(5)       & 0.39(4) \\
 $6_{4,3}-5_{4,2}$        & 141751.492  & 141751.9(10)        & 14.0(8)       & 0.21(2)$^a$\\
 $6_{4,2}-5_{4,1}$        & 141755.360  & 141755.5(10)        & 13.1(8)       & 0.19(2)$^a$\\
 $6_{2,4}-5_{2,3}$        & 145325.875  & 145326.3(10)        & 11.9(8)       & 0.41(4) \\
 $7_{0,7}-6_{0,6}$        & 158499.228  & 158499.6(5)         & 12.6(5)       & 0.63(6) \\
\hline
\multicolumn{5}{c}{V636\,Mon} \\
\hline
 $6_{2,5}-5_{2,4}$        & 140920.171  & 140921.2(10)       & 21.3(10)     & 0.35(7) \\
 $6_{4,3}-5_{4,2}$        & 141751.492  & 141751.8(4)        & 20.3(8)      & 0.17(3)$^a$\\
 $6_{4,2}-5_{4,1}$        & 141755.360  & 141754.4(3)        & 19.7(6)      & 0.23(5)$^a$ \\
 $6_{2,4}-5_{2,3}$        & 145325.875  & 145326.5(10)       & 23.3(10)     & 0.37(1) \\
 $7_{0,7}-6_{0,6}$        & 158499.228  & 158499.9(10)       & 21.8(10)     & 0.61(2) \\
\hline
\multicolumn{5}{c}{IRC\,+20370} \\
\hline
 $6_{2,5}-5_{2,4}$        & 140920.171 & 140920.2(1)      & 13.0(2)  & 0.98(9) \\
 $6_{4,3}-5_{4,2}$        & 141751.492 & 141751.6(5)      & 13.2(5)  & 0.53(5)$^a$ \\
 $6_{4,2}-5_{4,1}$        & 141755.360 & 141755.3(5)      & 13.4(5)  & 0.51(5)$^a$ \\
 $6_{2,4}-5_{2,3}$        & 145325.875 & 145325.9(3)      & 13.2(5)  & 1.12(11) \\
 $7_{0,7}-6_{0,6}$        & 158499.228 & 158499.3(2)      & 14.6(1)  & 1.83(18) \\
\hline
\multicolumn{5}{c}{W\,Ori} \\
\hline
 $6_{2,5}-5_{2,4}$        & 140920.171 & 140920.2(10)  & 9.4(10)     & 0.12(5)$^c$ \\
 $6_{4,3}-5_{4,2}$        & 141751.492 & 141751.5(10)  & 10.5(10)    & 0.06(1)$^c$ \\
 $6_{4,2}-5_{4,1}$        & 141755.360 & 141754.6(10)  & 10.9(10)    & 0.06(3)$^c$ \\
 $6_{2,4}-5_{2,3}$        & 145325.875 & 145325.2(10)  & 8.2(10)     & 0.11(2)$^c$ \\
 $7_{0,7}-6_{0,6}$        & 158499.228 & 158499.9(5)   & 12.1(10)    & 0.31(6)\\
\hline
\multicolumn{5}{c}{CRL\,67} \\
\hline
 $6_{2,5}-5_{2,4}$        & 140920.171 & 140920.3(1)   & 14.6(1)   & 1.57(1) \\
 $6_{4,3}-5_{4,2}$        & 141751.492 & 141751.5(5)   & 15.6(5)   & 0.72(7)$^a$ \\
 $6_{4,2}-5_{4,1}$        & 141755.360 & 141755.2(5)   & 14.9(5)   & 0.96(9)$^a$ \\
 $6_{2,4}-5_{2,3}$        & 145325.875 & 145325.9(2)   & 14.6(4)   & 1.61(2) \\
 $7_{0,7}-6_{0,6}$        & 158499.228 & 158499.3(3)   & 15.8(4)   & 2.47(2) \\
\hline
\multicolumn{5}{c}{CRL\,190} \\
\hline
 $6_{2,5}-5_{2,4}$        & 140920.171 & 140920.6(1)    & 16.4(1) & 1.08(10) \\
 $6_{4,3}-5_{4,2}$        & 141751.492 & 141751.6(3)    & 16.3(5) & 0.29(3)$^a$ \\
 $6_{4,2}-5_{4,1}$        & 141755.360 & 141755.1(2)    & 15.2(5) & 0.37(4)$^a$ \\
 $6_{2,4}-5_{2,3}$        & 145325.875 & 145325.9(1)    & 15.4(1) & 0.12(1) \\
 $7_{0,7}-6_{0,6}$        & 158499.228 & 158499.6(1)    & 16.8(4) & 0.77(7) \\
\hline
\multicolumn{5}{c}{S\,Aur} \\
\hline
 $6_{2,5}-5_{2,4}$  & 140920.171  & 140920.3(10)  & 23.7(10)  & 0.11(2)$^c$ \\
 $6_{4,3}-5_{4,2}$  & 141751.492  & 141753.1(10)  & 22.2(14)  & 0.13(3)$^{b, c}$ \\
 $6_{4,2}-5_{4,1}$  & 141755.360  & -  & -   & -  \\
 $6_{2,4}-5_{2,3}$  & 145325.875  & 145324.9(10)  & 25.7(15)  & 0.15(3)$^c$ \\
 $7_{0,7}-6_{0,6}$  & 158499.228  & 158499.1(5)   & 24.4(8)   & 0.28(6) \\
\hline
\multicolumn{5}{c}{V\,Aql} \\
\hline
 $6_{2,5}-5_{2,4}$  & 140920.171  & 140920.1(5)     & 6.7(4)        & 0.31(6) \\
 $6_{4,3}-5_{4,2}$  & 141751.492  & 141751.7(10)    & 8.0(10)       & 0.26(5)$^b$ \\
 $6_{4,2}-5_{4,1}$  & 141755.360  & 141755.3(10)    & 9.7(10)       & 0.12(2)$^b$\\
 $6_{2,4}-5_{2,3}$  & 145325.875  & 145326.1(5)     & 8.1(6)        & 0.38(7) \\
 $7_{0,7}-6_{0,6}$  & 158499.228  & 158499.2(5)     & 8.0(5)        & 0.60(12) \\
\hline
\multicolumn{5}{c}{CRL\,2513} \\
\hline
 $6_{2,5}-5_{2,4}$        & 140920.171  & 140922.1(10)   & 20.3(10)    & 0.22(4) \\
 $6_{4,3}-5_{4,2}$        & 141751.492  & 141751.7(10)   & 20.6(10)    & 0.23(4)$^b$ \\
 $6_{4,2}-5_{4,1}$        & 141755.360  & -              & -           &  - \\
 $6_{2,4}-5_{2,3}$        & 145325.875  & 145326.3(10)   & 25.6(10)    & 0.29(6)$^c$ \\
 $7_{0,7}-6_{0,6}$        & 158499.228  & 158499.8(5)   & 26.9(4)      & 0.58(11)\\
\hline
\multicolumn{5}{c}{CRL\,2477} \\
\hline
 $6_{2,5}-5_{2,4}$        & 140920.171  & 140920.1(1)      & 18.5(10)  & 0.46(4) \\
 $6_{4,3}-5_{4,2}$        & 141751.492  & 141751.5(2)      & 18.2(4)  & 0.28(6)$^a$ \\
 $6_{4,2}-5_{4,1}$        & 141755.360  & 141755.2(2)      & 19.9(4)  & 0.30(6)$^a$\\
 $6_{2,4}-5_{2,3}$        & 145325.875  & 145325.7(5)      & 19.3(5)  & 0.51(1) \\
 $7_{0,7}-6_{0,6}$        & 158499.228  & 158499.5(5)      & 20.0(5)  & 0.72(14) \\
\hline
\multicolumn{5}{c}{CRL\,2494} \\
\hline 
 $6_{2,5}-5_{2,4}$                & 140920.171  & 140920.2(5)   & 17.4(8)  & 1.41(1) \\
 $6_{4,3}-5_{4,2}$                & 141751.492  & 141751.8(5)   & 18.1(5)  & 0.81(1)$^a$ \\
 $6_{4,2}-5_{4,1}$                & 141755.360  & 141755.3(5)   & 18.1(5)  & 0.91(1)$^a$\\
 $6_{2,4}-5_{2,3}$                & 145325.875  & 145326.7(5)   & 18.3(8)  & 1.53(1) \\
 $7_{0,7}-6_{0,6}$                & 158499.228  & 158499.5(3)   & 19.5(4)  & 2.10(2) \\
\hline
\multicolumn{5}{c}{RV\,Aqr} \\
\hline
 $6_{2,5}-5_{2,4}$          & 140920.171  & 140920.7(10)     & 14.3(10)    & 0.37(7) \\
 $6_{4,3}-5_{4,2}$          & 141751.492  & 141751.4(5)      & 14.7(10)    & 0.18(3)$^a$\\
 $6_{4,2}-5_{4,1}$          & 141755.360  & 141755.3(5)      & 14.2(10)    & 0.18(3)$^a$\\
 $6_{2,4}-5_{2,3}$          & 145325.875  & 145325.9(10)     & 15.6(10)    & 0.39(8) \\
 $7_{0,7}-6_{0,6}$          & 158499.228  & 158499.6(5)      & 15.5(5)     & 0.65(10) \\
\hline
\end{longtable}
\tablefoot{
Numbers in parentheses are 1$\sigma$ uncertainties in units of the last digits.\\
$^a$ Blend of two lines, each of which could be fitted individually. \\
$^b$ Blend of two lines which could not be fitted individually. Only one component was fitted. \\
$^c$ Marginal detection.}
}

\longtab{
\begin{longtable}{lclll}
\caption{Observed parameters of the SiC $^3\Pi_2$ $J=4-3$ line} \label{table:sic_lines} \\
\hline \hline
\multicolumn{1}{l}{Source} & \multicolumn{1}{c}{$\nu_{calc}$} & \multicolumn{1}{c}{$\nu_{obs}$} & \multicolumn{1}{c}{$V_{e}$}      & \multicolumn{1}{l}{$\int T_A^* dv$} \\
                                       & \multicolumn{1}{c}{(MHz)}           & \multicolumn{1}{c}{(MHz)}          & \multicolumn{1}{l}{(km s$^{-1}$)} & \multicolumn{1}{l}{(K km s$^{-1}$)} \\
\hline
IRC\,+10216  & 157494.101 & 157494.1(1)   & 15.4(8)   & 5.86(6) \\
CIT\,6             & 157494.101 & 157493.2(10)  & 15.4(8)   & 0.96(9) \\
CRL\,3068     & 157494.101 & 157494.2(5)   & 15.7(5)   & 0.29(6) \\
IRC\,+30374  & 157494.101 & 157495.5(10)  & 25.1(10)  & 0.18(3)$^a$ \\
LP\,And         & 157494.101 & 157493.8(5)   & 14.4(4)   & 0.56(6) \\
V\,Cyg           & 157494.101 & 157496.3(10)  & 11.3(10)  & 0.11(2)$^a$ \\
V384\,Per      & 157494.101 & 157493.7(10)  & 15.2(10)  & 0.17(3)$^a$ \\
IRC\,+60144  & 157494.101 & 157492.8(10)  & 20.2(10)   & 0.12(6)$^a$ \\
IRC\,+20370  & 157494.101 & 157494.3(10)  & 15.8(10)   & 0.13(2)$^a$ \\
CRL\,67         & 157494.101 & 157494.8(10)  & 16.6(10)   & 0.28(6)$^a$\\CRL\,2477    & 157494.101 & 157493.7(10)  & 18.7(10)    & 0.10(2)$^a$ \\
CRL\,2494    & 157494.101 & 157493.5(10)  & 21.8(10)    & 0.23(4)$^a$\\
\hline
\hline
\end{longtable}
\tablefoot{
Numbers in parentheses are 1$\sigma$ uncertainties in units of the last digits.\\
$^a$ Marginal detection.}
}

\end{appendices}

\end{document}